# Photon Detection Using Cerium Bromide Scintillation Crystals

by
Mohamad Akkawi
Supervisor: Dr. Kevin Fissum
Co-Supervisors: Dr. Francesco Messi, Dr. Hanno Perrey

Division of Nuclear Physics
Department of Physics
Lund University

FYS016

LUND UNIVERSITY

# Abstract

The project involved the development of a precision radioactive Source-Positioning System to enable the absolute characterization of a gamma-ray Detector-Development Platform at the Division of Nuclear Physics at Lund University in conjunction with the Detector Group of the European Spallation Source. The gamma-ray radiation associated with an actinide-Beryllium source was mapped using a cerium bromide scintillation crystal coupled to photomultiplier tube at various relative source and detector configurations. The data obtained are used to determine the spatial and energy profiles of the neutron/gamma-ray beams associated with the source. This will in turn enable the unfolding of detector responses via simulation toolkits such as Geant4 and may also facilitate the future design of a boron-carbide beam-hardening sleeve.

# Abbreviations

| | |
|---|---|
| CAD | Computer-Aided Design |
| $CeBr_3$ | Cerium Bromide |
| DDP | Detector-Development Platform |
| ESS | European Spallation Source |
| FWHM | Full-Width-at-Half-Maximum |
| MCA | Multi-Channel Analyzer |
| PID | Proportional–Integral–Derivative |
| PMMA | Poly-Methyl-Methacrylate |
| PMT | Photomultiplier Tube |
| SPS | Source-Positioning System |
| STF | Source-Testing Facility |

# Contents

# List of Figures

# List of Tables

# Chapter 1: Background

## 1.1 Nuclear Physics

### 1.1.1 Binding Energy and Mass Defect

Binding energy is a concept used to quantify the amount of energy required to hold an atomic nucleus together. The binding energy exists due to the attractive force within the nucleus, called the strong nuclear force [1].

Another representation of the binding energy is called the mass defect. The mass of the nucleus of any atom containing more than one nucleon will always be less than the combined mass of the individual nucleons. This difference in mass is referred to as the mass defect and is proportional to the binding energy according to Einstein's mass-energy equivalence equation [2].

The binding energy per nucleon is thus equal to the amount of energy required to separate one nucleon from the nucleus. This energy can therefore be used to determine the stability of an atom. The higher the binding energy per nucleon, the more stable the atom [2].

As an example, the binding energy of $^{241}$Am is 7.54 MeV per nucleon (see Appendix A).

### 1.1.2 Q-value and Alpha Decay

The Q-value is a related concept defined as the mass difference between the initial and final states of a system. It represents the energy released in or required by a nuclear reaction. If released, this energy is shared kinetically by the particles in the final state [3].

Alpha decay (α-decay) is a type of radioactive decay where the mother nucleus emits an alpha particle (α-particle). An α-particle is another name for the nucleus of a $^{4}$He atom, which means that the mother nucleus will have its atomic mass reduced by 4 and its atomic number reduced by 2 after emitting an α-particle, producing a so-called daughter nucleus [4].

The ground state Q-value for the α-decay of $^{241}$Am decay is 5.634 MeV (see Appendix B).

### 1.1.3 Variation of Binding Energy per Nucleon

In general, elements with low atomic mass may undergo nuclear fusion, where multiple light atomic nuclei are fused together under specific conditions to form a final product that has a greater binding energy per nucleon. Elements with high atomic mass may undergo nuclear fission, where heavy atomic nuclei are split into multiple daughter nuclei with greater binding energy per nucleon. These nuclear reactions emit various types of radiation, including (but not limited) to: α-particles, free nucleons, and electromagnetic radiation [2]. Both types of reactions produce daughter nuclei/final-state nuclei with higher binding energies per nucleon as this is a measure of isotopic stability. This is further illustrated in Figure 1.1 below.

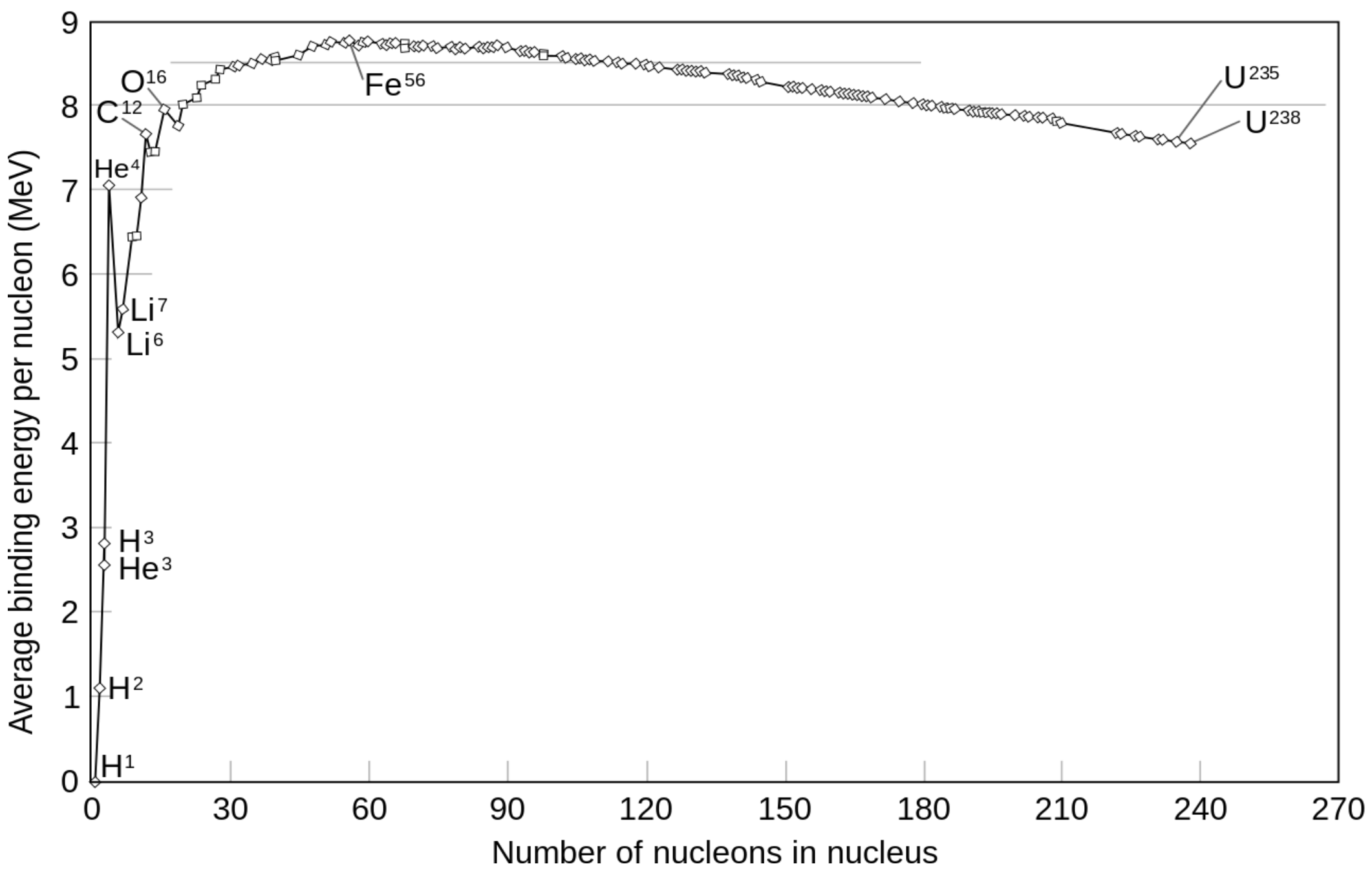

Figure 1.1: The nuclear binding energy per nucleon versus the number of nucleons. The isotopes to the left of $^{56}$Fe undergo fusion to produce more stable products, whereas isotopes to the right undergo fission. Figure from Ref. [5].

## 1.2 Photon Interaction with Matter

### 1.2.1 Photons

Photons can be thought of as particles or waves. They have a constant (light) speed $c$ with a varying wavelength depending on their energy. Photons with high energies (over 100 keV) are called gamma-rays (γ-rays). They are ionizing and interact with matter with a certain probability described by the interaction "cross section". Cross sections greatly depend on photon energy [6].

### 1.2.2 Photoelectric Effect

When a photon interacts with an atom and then completely disappears, this is known as the photoelectric effect. A photoelectron characterized by the energy of the incident photon is then emitted by the atom from one of its bound shells. The vacancy is then immediately filled by a free electron from the rearrangement of electrons in the atomic shells. The electron rearrangement then releases one or more photons [6].

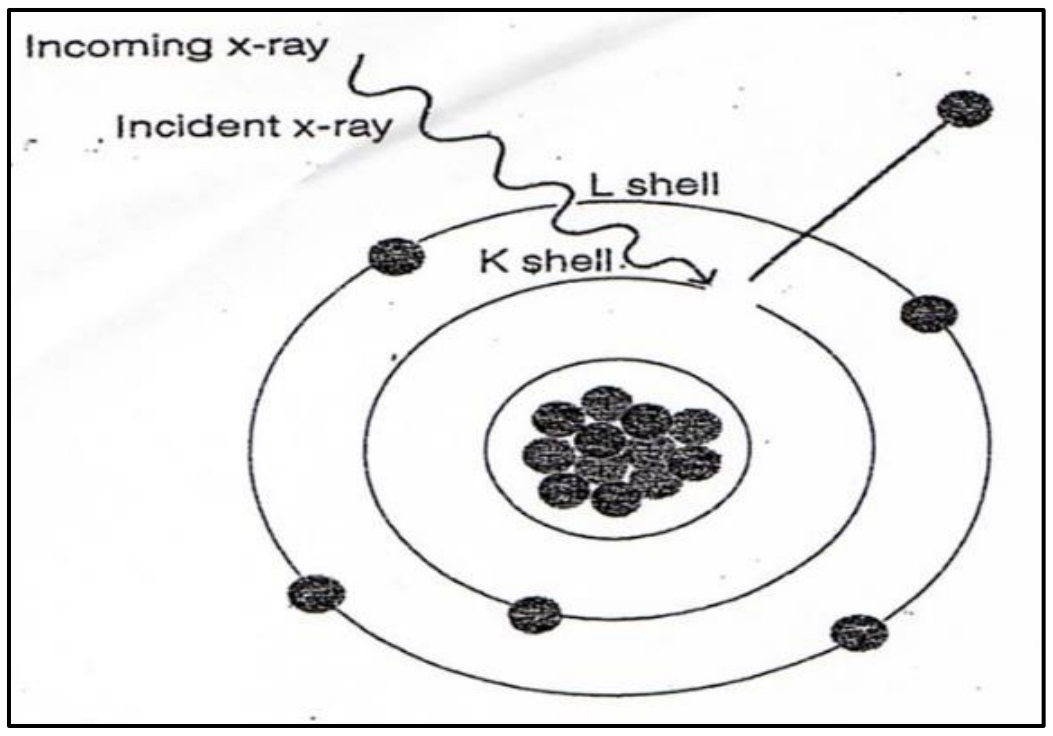

Figure 1.2: The photoelectric effect due to an X-ray interaction. Figure from Ref. [7].

The photoelectron is emitted with an energy given by:

$$E_e = hv - E_b$$

where $E_b$ represents the binding energy of the photoelectron in its original shell and $hv$ is the original photon energy. $E_b$ becomes less significant for high-energy incident photons i.e γ -rays [6].

### 1.2.3 Compton Scattering

Compton scattering occurs when there is a collision between a photon and a "resting" electron in the absorbing material. Only a portion of the photon energy is transferred to the then-freed electron. The probability of Compton scattering increases linearly with the atomic number Z of the absorber material due to an increase in the number of electrons available as scattering targets [6].

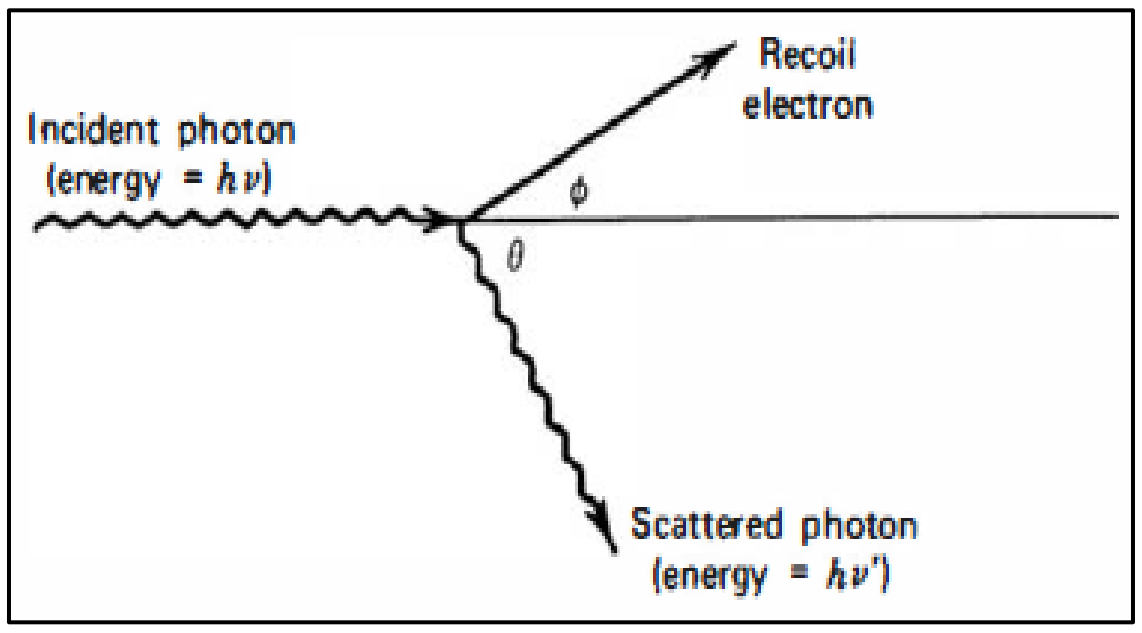

Figure 1.3: Compton Scattering. Figure from Ref. [6].

The scattered photon energy is given by:

$$hv' = \frac{hv}{1 + \frac{hv}{m_0c^2}(1 - cos\theta)}$$

where $m_0c^2$ is the rest-mass energy of the electron at 0.511 MeV, $hv$ is the orginal photon energy, and $\theta$ is the scattering angle.

The energy transferred to the electron increases as the photon scattering angle increases. However, some of the original energy is always retained by the incident photon, even in the extreme head-on collision case of 180° [6].

### 1.2.4 Pair Production

When a photon disappears during an interaction and is replaced by an electron-positron pair, pair production has occurred. The phenomenon is only possible when the γ-ray has at least twice the rest-mass energy of an electron i.e 1.022 MeV. Any extra photon energy above the minimum required to create the pair is shared by the electron-positron pair as kinetic energy.

The positron will eventually be annihilated by another electron in the absorbing material. The annihilation results in the production of two new photons with 0.511 MeV each.

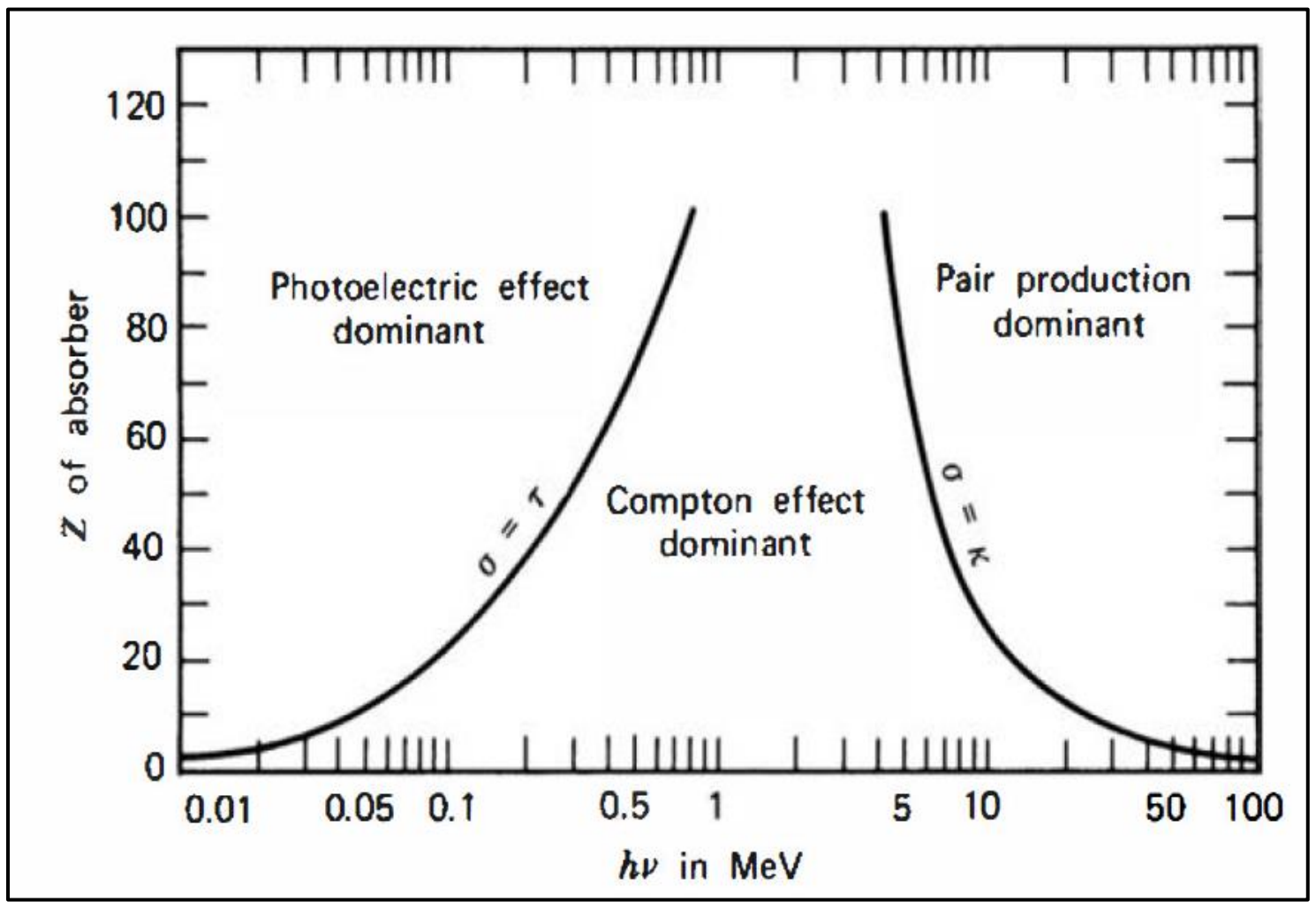

Figure 1.4: Interaction Dominance Z versus photon energy in MeV. Figure from Ref. [6].

The significance of the main interactions described above for different absorber materials and photon energies is shown in Figure 1.4 above. Pair production is the most likely form of γ-ray interaction when the incident photon energy exceeds 5 MeV.

# Chapter 2: Materials and Methods

## 2.1 Facility Apparatus

### 2.1.1 Source Tank

A Source-Testing Facility (STF) has been established at the Division of Nuclear Physics at Lund University. The main purposes of the STF are the characterization of neutron and γ-ray detectors and the conducting of shielding tests.

The STF has a water-filled tank shown in Figure 2.1 below which is used for shielding and creating beams from the source. Up to four characterization experiments can be conducted simultaneously using the cylindrical beam ports located on each face of the tank.

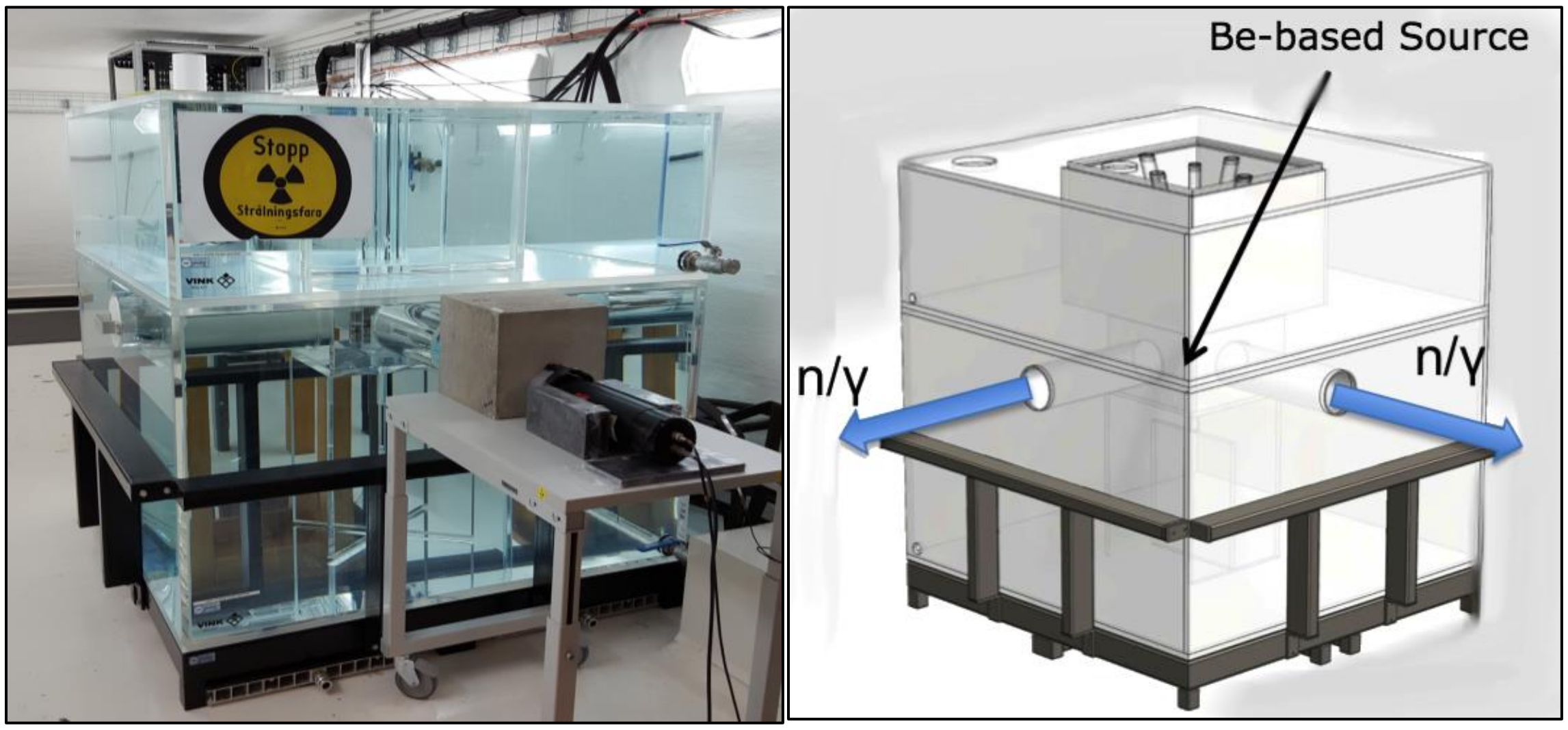

Figure 2.1: The Source Tank at the Source-Testing Facility. Left) Photograph of the tank at the STF (by author). Dimensions are 140 cm X 140 cm X 140 cm. The tank is made from Plexiglas and filled with ultra-pure deionized water. Right) Ideal $n/\gamma$ beams from a Be-based source are illustrated. Beam port dimensions are 172 mm (diameter) by 500 mm (depth). Figure from Ref. [8].

### 2.1.2 Shielding and Source Position

Shielding by the tank is a function of geometry and hence all four $n/\gamma$ beams are expected to be similar due to the symmetry of the tank. The geometry of the tank also ensures that the radiation dose to personnel outside the tank is always less than 0.5 μSv/h when used with a Be-based source [9].

Figure 2.2 illustrates key vertical source positions across the center of the tank relative to the STF floor. The vertical position of the beam ports is referred to as "production" position.

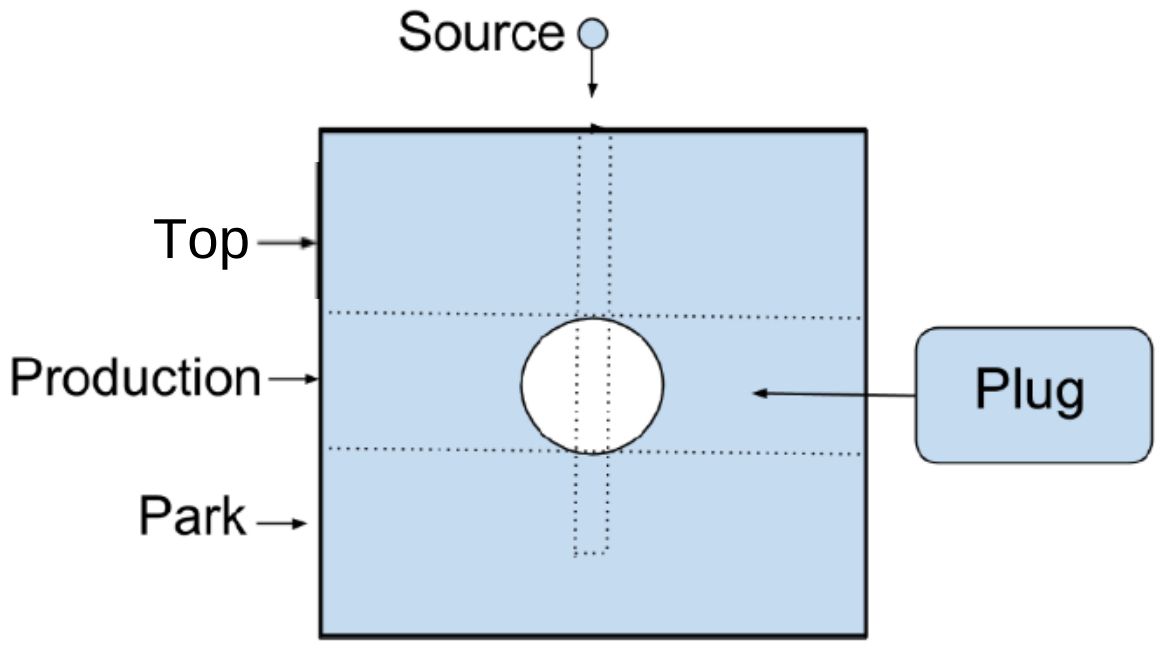

Figure 2.2: Drawing of the Source Tank. Relative to the STF floor: the top position is at 131 cm, the production position is at 110 cm and the park position is at 81 cm. A fixed vertical Plexiglas tube (26 mm in diameter) is used to guide the source through the center of the tank. Figure from Ref. [10].

Positioning the radioactive source at the production position will result in the most intense radioactive beams due to the lack of shielding. The park position is ideally an "off-switch" for the source-associated radiation. A polyethylene plug is also used to seal a beam port when it is not in use [9].

### 2.1.3 Source-Positioning System

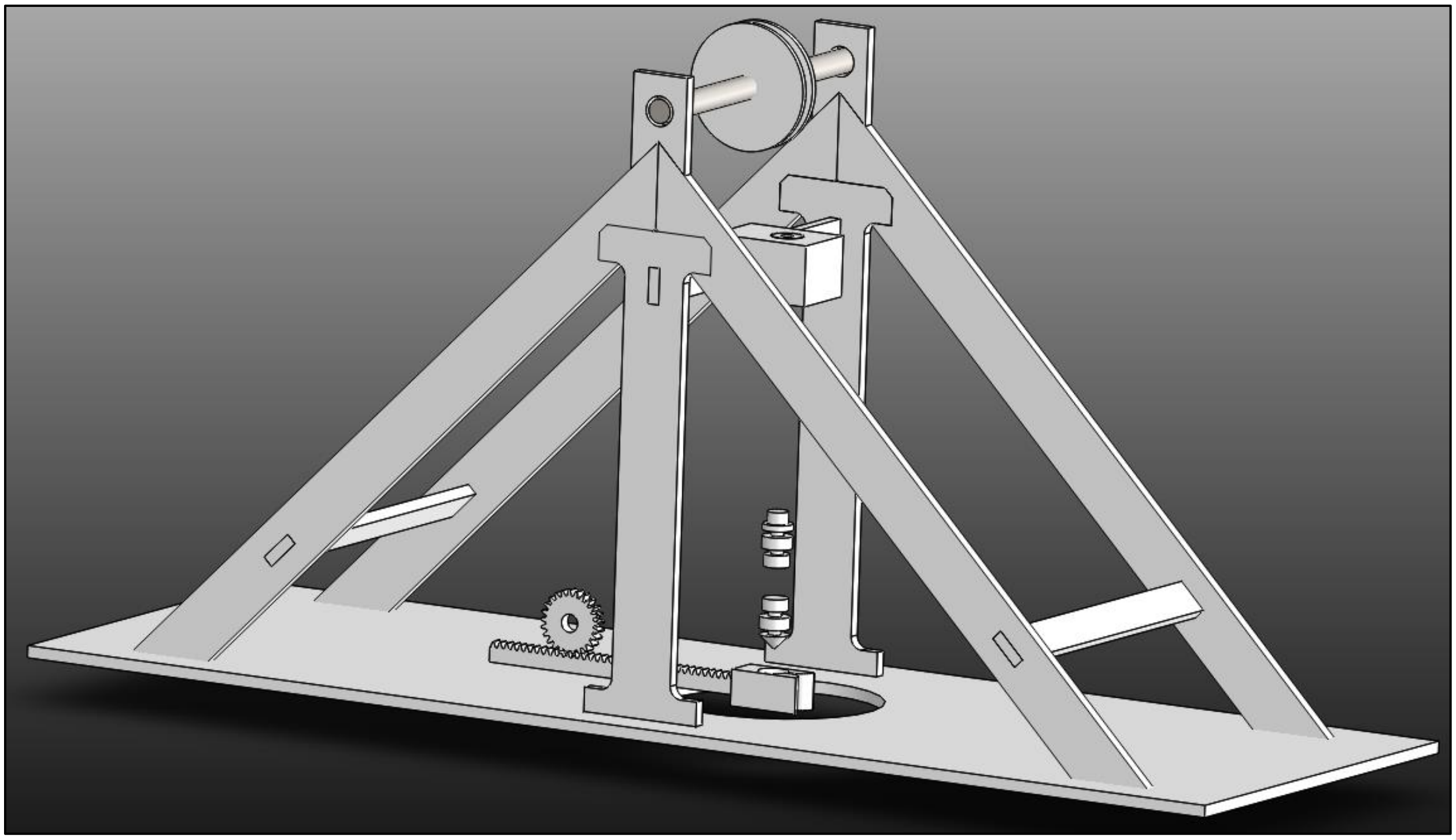

Figure 2.3: Computer Aided Design Drawing of Source-Positioning System (by author).

The primary function of the Source Positioning System (SPS) shown in Figure 2.3 above is to move a radioactive source vertically across the production tubes of the tank. The secondary function of the SPS is to reposition a source with enough precision such that the originally measured spectrum can be reproduced.

The SPS uses a stepper motor and Hall Effect sensor to accomplish the primary and secondary functions. The Hall Effect sensor is fixed to the critical source position i.e the production position. Once the magnet attached to the source elevator is detected, the stepper motor stops. Using a negative unity feedback loop with a PID controller, the system then continuously corrects for any overstepping or disturbances until the steady state error is less than 1%. Control of the SPS was achieved with a Raspberry Pi3 and mounted Gertbot driver. A list of SPS components can be found in Appendix D.

The SPS was built using Plexiglas and laser cut for extra precision. The elevator and gate were 3D printed. The function of the elevator is to move and hold the source in place. The elevator is moved using a pulley and fishing wire and the source is secured inside it by the tension of the fishing wires surrounding it which is due to the weight of the source.

The gate acts as a resting place for the elevator. When the elevator is placed onto the gate, the tension in the fishing wires is released which enables the user to remove/replace the source. The gate is moved by a rack and pinion which is also controlled by a stepper motor. The secondary function of the gate is to physically prevent the source from being ejected from the tank in the case of a system failure or user error.

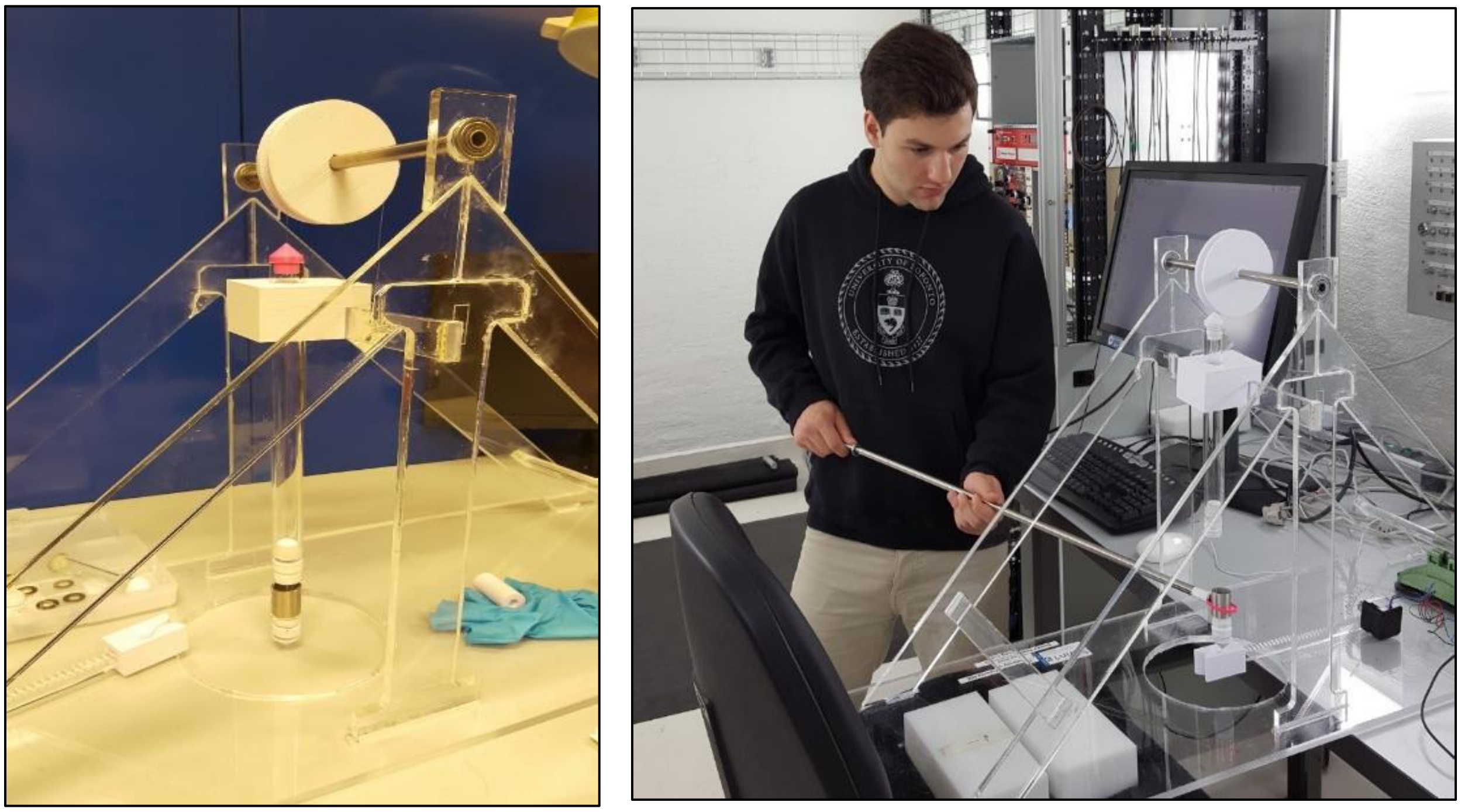

Figure 2.4: Source-Positioning System Demonstration. Left) Dummy source inside elevator. Photograph by author. Right) Dummy source being removed from elevator. Photograph by Emil Rofors.

## 2.2 Be-based Neutron/Gamma Source

### 2.2.1 Americium-241

$^{241}$Am is an element that is no longer found in nature and is instead synthesized through a lengthy and costly process. It is a radioactive metal with and an atomic number of 95 and a half-life of 432.2 years [11]. It decays via five individual α emissions to become $^{237}$Np. The dominant decay mechanism is the one that decays to $^{237}$Np from the intermediate state at 60 keV as shown in Figure 2.5 below. Energy is released via the α-particle depending on the post-decay state of $^{237}$Np. Almost instantaneously afterwards, a γ-ray is released when $^{237}$Np decays to its ground state [9].

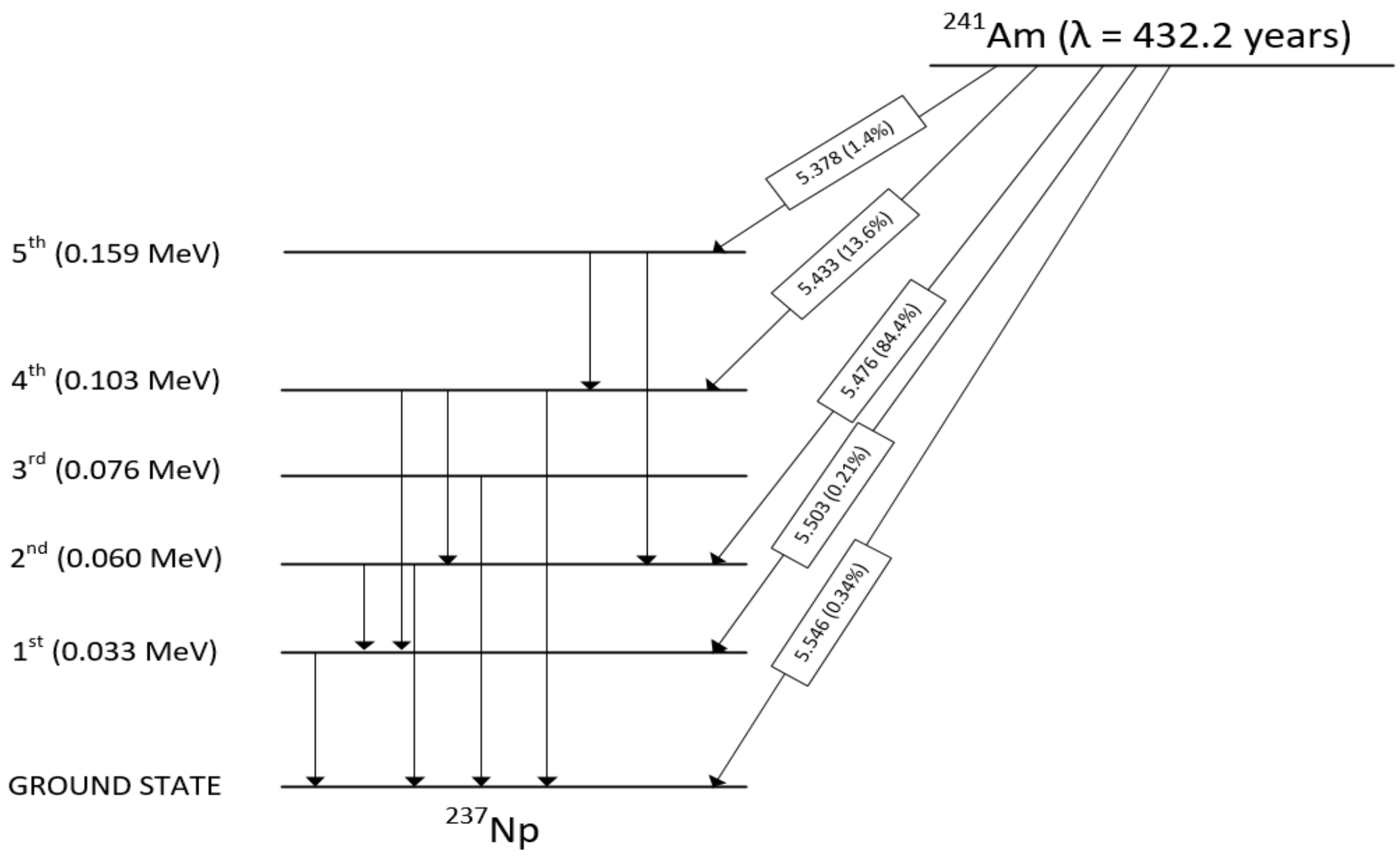

Figure 2.5: The various α-decay Pathways of $^{241}$Am. The most common energy is 60 keV. Figure from Ref. [1].

### 2.2.2 Free-Neutron Production

When an α-particle is emitted from the decay of an actinide (such as $^{241}$Am) and collides with $^{9}$Be, a free neutron as well as a $^{12}$C may be created. The recoiling $^{12}$C can be in either its ground state or one of two excited states.

$$\alpha + {}^{9}_{4}Be \rightarrow {}^{12}_{6}C + n$$

The first excited state of $^{12}$C will decay to its ground state almost immediately by emitting a γ-ray with energy 4.44 MeV [9]. The Q-value of the reaction to the ground state is 5.7 MeV (see Appendix C).

### 2.2.3 Energy Distribution of AmBe Emissions

From an AmBe source, the total energy that can be shared by all the reaction products is about 11.3 MeV. Hence this is the maximum free neutron energy. A free neutron with energy less than 11.3 MeV implies that some energy was lost to intermediate processes or the kinetic energy of the recoiling $^{12}C$. The most common state of the recoiling $^{12}C$ is the first excited state at 55%. On average, the corresponding neutron will have an energy of 4.5 MeV [9].

### 2.2.4 Source Geometry and Housing

The AmBe mixture is encapsulated in a stainless-steel capsule and emits γ-rays and free neutrons roughly isotropically. The neutron emission rate is $(1.106 \pm 0.015) \times 10^6$ neutrons/second.

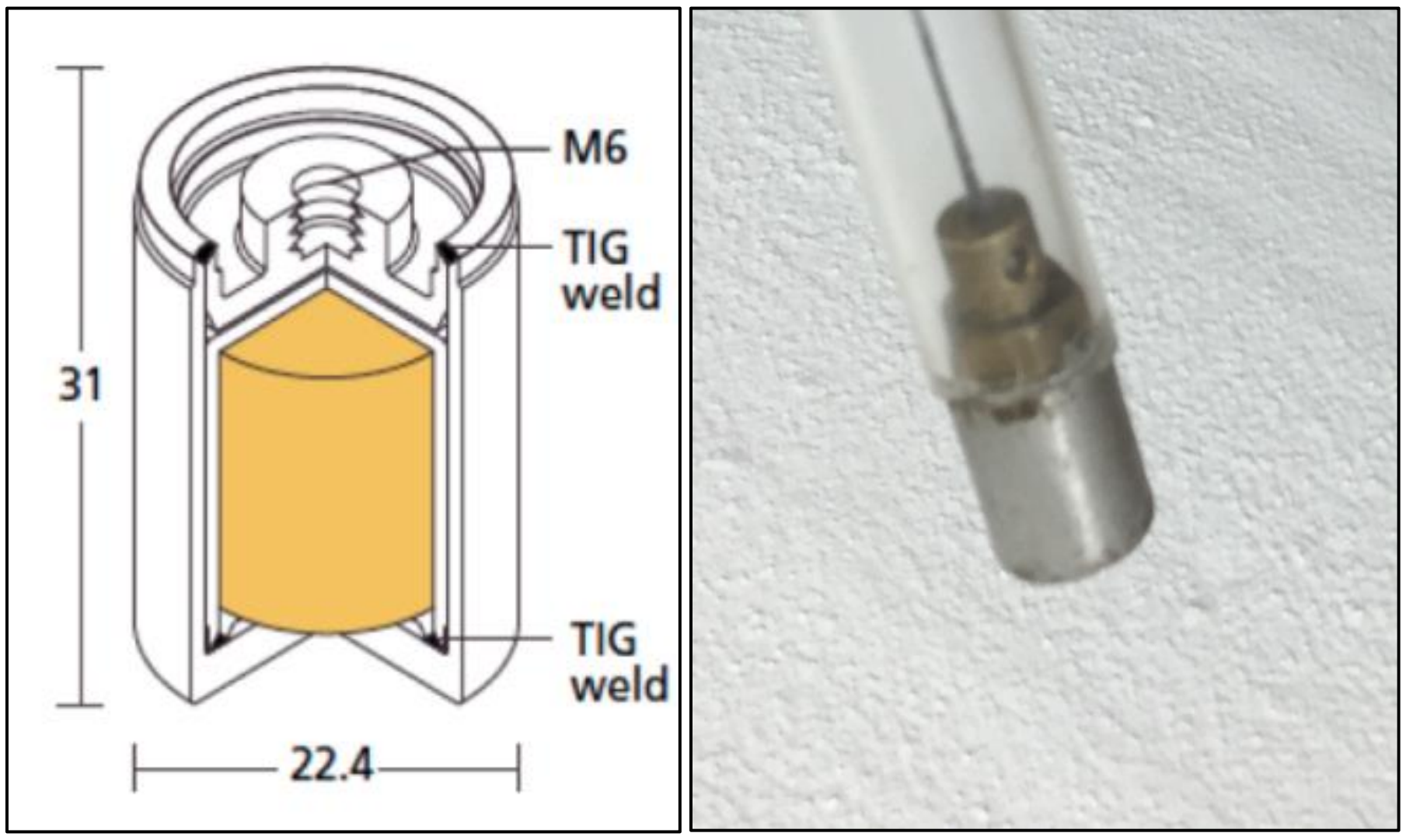

Figure 2.6: AmBe Source. Left) Schematic drawing (dimensions are given in mm). The yellow volume is the radioactive material. Drawing from Ref. [12]. Right) Photograph of source from Ref. [10].

## 2.3 $CeBr_3$ Gamma-ray Detector

### 2.3.1 Overview

The main function of a γ-ray detector is to convert incident γ-rays into an electrical pulse. The $CeBr_3$ detector used to characterize the AmBe-associated γ-ray beam exiting the beam port is made up of two integrated modules (B style) consisting of a Photomultiplier Tube (PMT) and a $CeBr_3$ inorganic crystal [13]. See Section 1.2 for a discussion of how the incident γ-rays interact with the $CeBr_3$ crystal.

After scintillation by the $CeBr_3$ crystal, the purpose of the PMT is to convert light signals into a current pulse. The photocathode and electron multiplier are the main components of the PMT. When a photon strikes the photocathode, electrons are emitted due to the photoelectric effect. The emitted photoelectrons are then multiplied by the first dynode through the last dynode as shown in Figure 2.7 which leads to an electron amplification of up to 1000000. Electrons finally arrive at the anode which collects them and outputs the electron current pulse to an external circuit [14].

The outer housing of the PMT maintains vacuum conditions inside the tube such that low-energy electrons can be beamed efficiently onto the first dynode by the internal electric fields of the focusing electrodes. The electric field is produced due to the electrode configuration and applied voltage. The maximum value for the applied voltage determines the maximum gain obtainable from the tube [15].

### 2.3.2 Nonlinearity

After each secondary emission, a cloud of positive ions is created which slowly disperses as it moves toward the cathode. High concentrations of these ions represent a space charge that can significantly alter the shape of the electric field within the detector. The electric field then affects the trajectories of electrons and causes some to be lost that would otherwise be collected. When the applied voltage too high, this space-charge effect becomes dominant in the pulse history which would introduce nonlinear effects [14]. The resistivity of the photosensitive layer in the PMT also limits the maximum allowable photocurrent. For Biakali layers, the limit is $2.5\ \frac{nA}{cm^2}$. Therefore the R6231 PMT used in this experiment has a current limit of roughly $41.55\ nA$.

High mean-anode currents can also reduce PMT lifetime. The maximum mean-anode current should be less than 100 μA and more stable performance is achieved by operating below 10 μA. With the correct voltage-divider design, all multiplier structures are very linear up to anode currents of 100 μA [14] (see Appendix E).

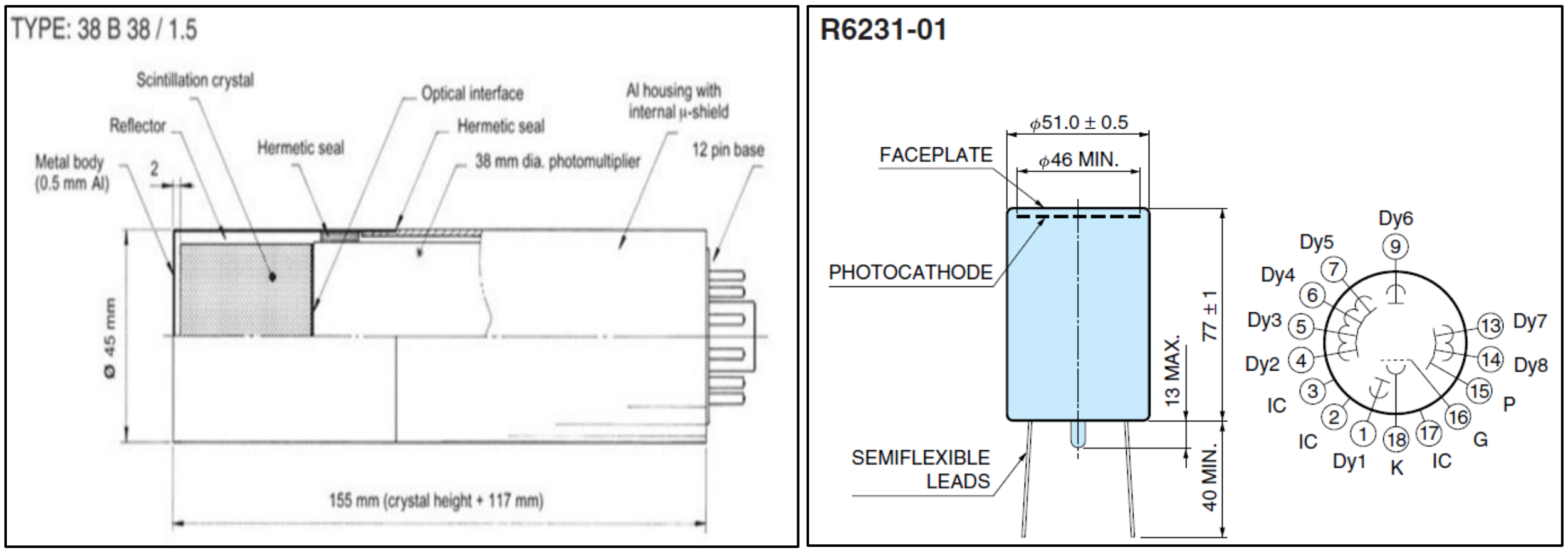

Figure 2.7: Photomultiplier tube Schematics. Left) Example of standard B-style detector with 38 mm PMT. Figure from Ref. [13]. Right) PMT model used with the $CeBr_3$ crystal. Illustration of its dynode configuration. Figure from Ref. [16].

### 2.3.3 Energy Resolution

A detector with an infinite energy resolution would measure delta-function peaks when operated in spectroscopic mode. However, as the resolution decreases, the energy of the incident γ-ray is deposited statistically and takes the form of a Gaussian distribution. Some counts will be detected above the incident energy and some below. It is standard to use the Full-Width-at-Half-Maximum (FWHM) of the normally distributed peak when determining the energy resolution of a detector. The FWHM is the width of the distribution at half its height. Typically, a Gaussian fit is conducted to approximate the peak energy and standard deviation which is used to calculate the FWHM. Typically, as FWHM increases resolution decreases [10].

The energy resolution is determined by

$$R = \frac{\Delta E_{FWHM}}{E}$$

where $\Delta E_{FWHM}$ is the FWHM in units of energy, and $E$ is the energy of the peak.

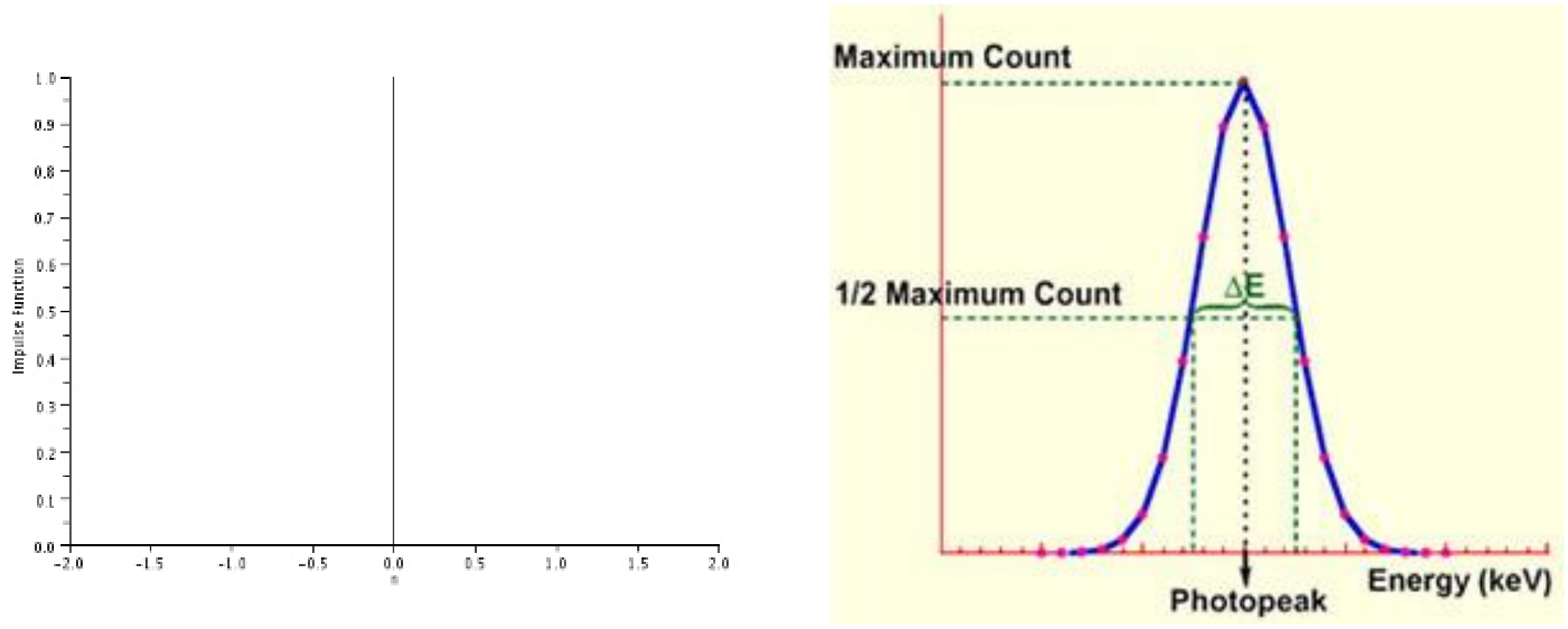

Figure 2.8: Photopeak Resolution. Left) photo peak as a delta function corresponding to a detector with infinite resolution. Figure from Ref. [17]. Right) Corresponding distribution skewed by detector resolution appears as a Gaussian function. Figure from Ref. [18].

# Chapter 3: Data Acquisition, Results and Discussion

## 3.1 Experimental Setup

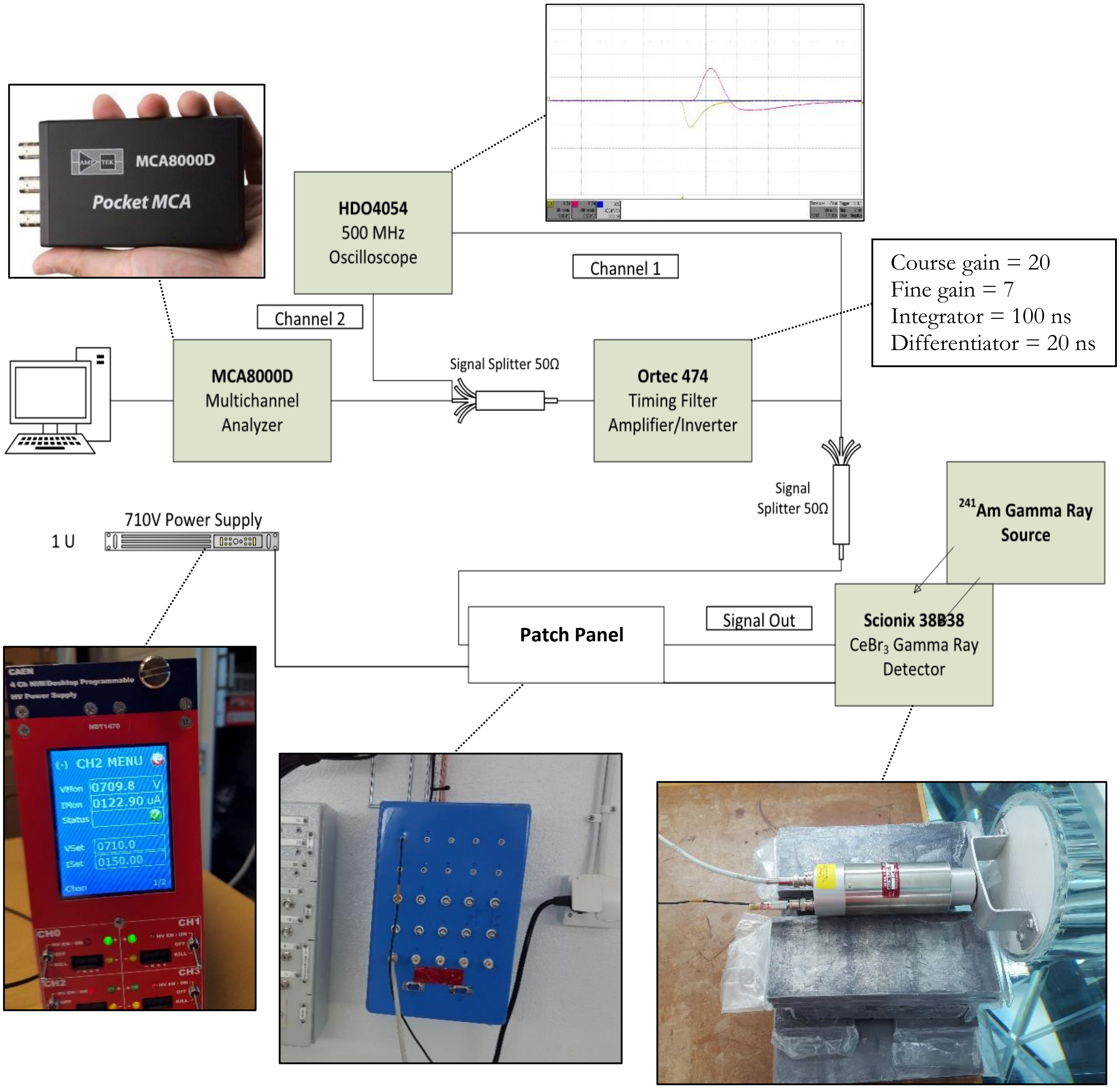

Figure 3.1: Schematic of $CeBr_3$ detector setup. Photographs by author.

### 3.1.1 Calibration and Multichannel Analyzers

The purpose of a multichannel analyzer is to convert an analog pulse amplitude to a specific discrete channel number. Ideally, an MCA output should result in a linear relationship between the pulse height and channel number. By adjusting the amplifier gain, we can spread the range of input pulses across the active range of the MCA as shown in Figure 3.2 [6].

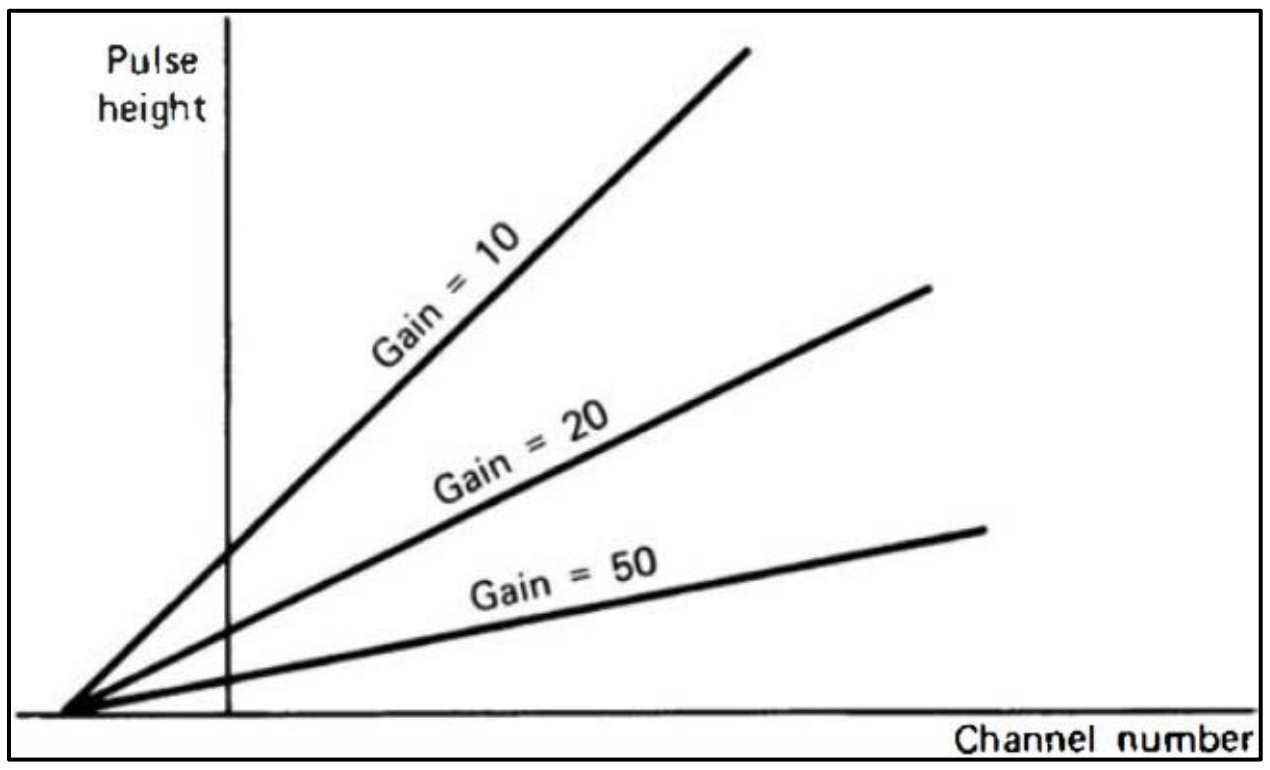

Figure 3.2: Gain vs. Channel Occupancy. Figure from Ref. [6].

Known sources of γ-rays are placed in front of a detector to record the channel numbers where the energy peaks occur on the spectrum. A linear fit is then performed to determine the energy calibration of the detector. After determining the slope and intercept of this linear relation, we are able to determine the energy of every MCA channel. Nonlinearity in high quality MCAs is usually less than 0.1 % [6].

## 3.2 Spectrometry Background

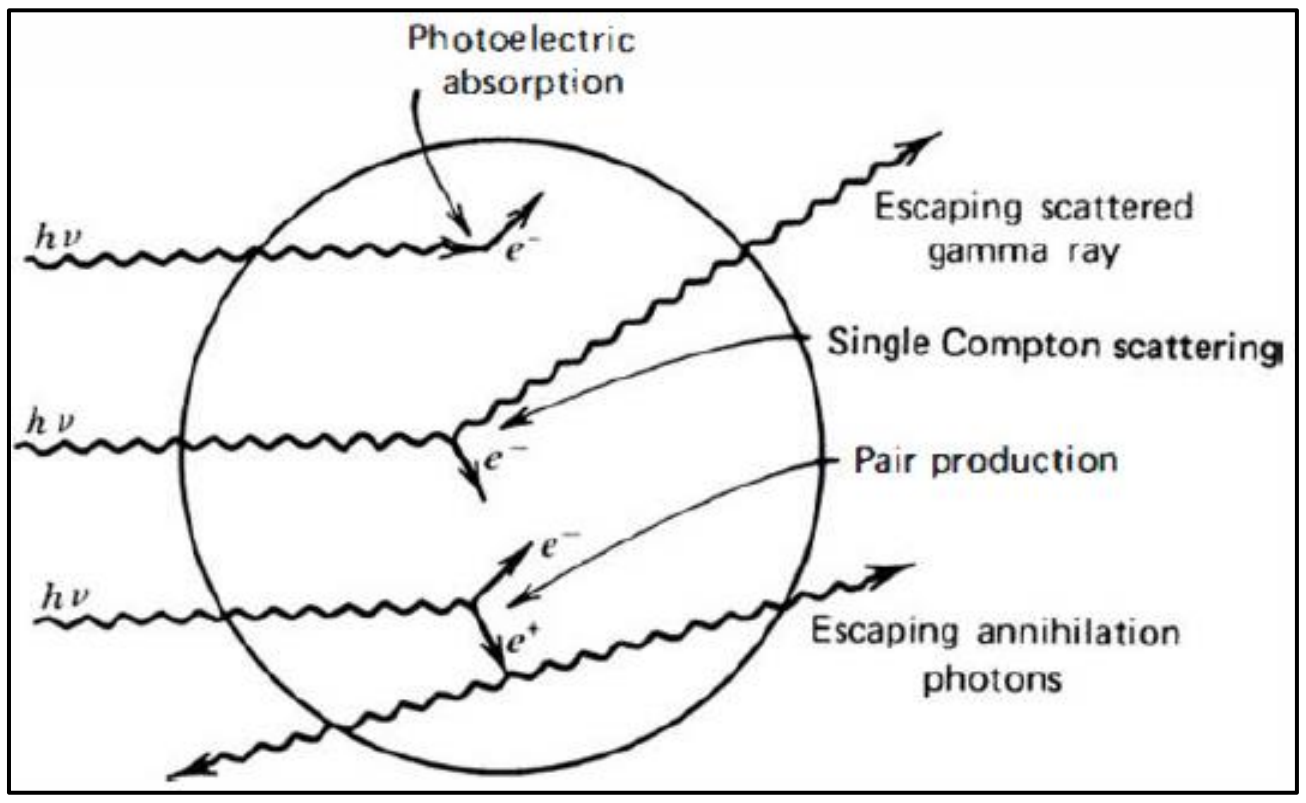

Figure 3.3: Variations in photon and matter interactions. Figure from Ref. [6].

### 3.2.1 Compton Edge

The extreme case of a 180° scatter of a γ-ray in a head-on collision with an electron represents the maximum energy that can be transferred to an electron in a single Compton interaction and is known as the Compton edge [6]. The maximum energy an electron can absorb in a single Compton scatter is given by

$$E_{max} = \frac{2E_\gamma^2}{m_0c^2 + eE_\gamma}$$

$E_{max}$ is the Compton-edge energy, $E_\gamma$ is the energy of the incident γ-ray, and $m_0c^2$ is the energy of an electron at rest.

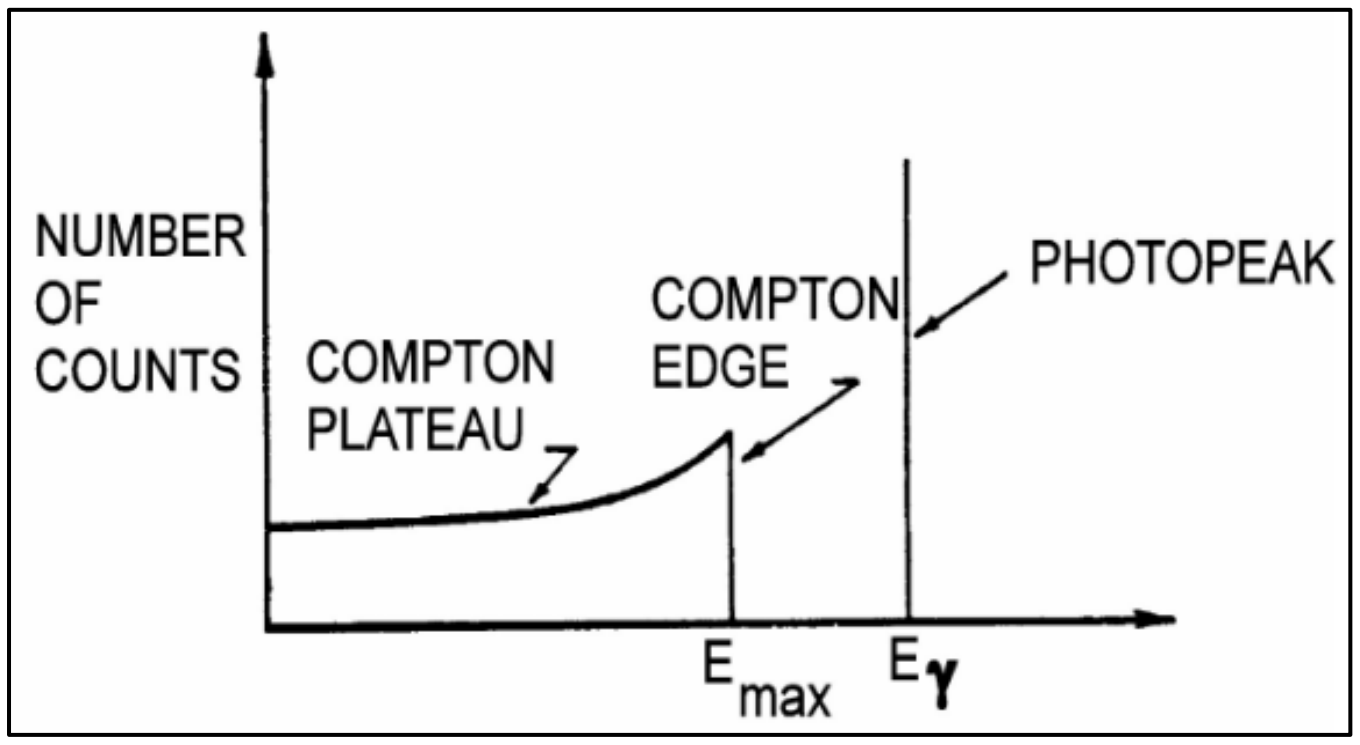

Figure 3.4: Spectrum of an ideal γ-ray detector with an infinite energy resolution. The photopeak is the result of the Photoelectric Effect. The Compton plateau is the range of Compton scatters which occurred at a lower incident angle than 180°. However, due to a finite energy resolution of any detector, the Compton edge will always be normally distributed rather than discrete [19]. Figure from Ref. [6].

### 3.2.2 First- and Second-Escape Peaks

The first- and second-escape peaks are directly related to the pair production phenomena explained in Section 1.2.5. When pair production occurs after an incident γ-ray over 1022 keV is absorbed, one or both the annihilation γ-rays may escape the detector. Hence, the histogram will be missing 511 keV or 1022 keV of the incident γ-ray energy [20]. The likelihood of this escape is a function of detector geometry. In general, the smaller the detector the greater the probability of escapes [6].

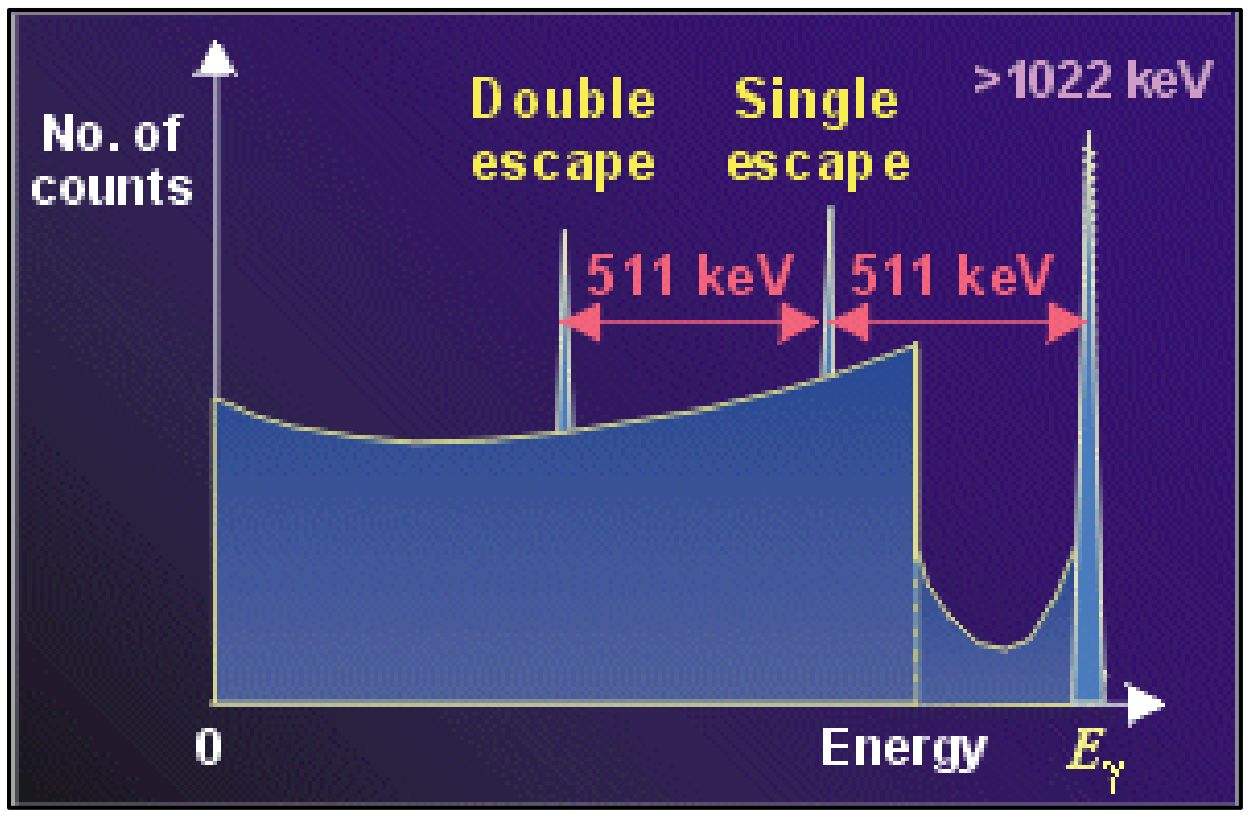

Figure 3.5: Ideal spectrum of a medium-sized detector with a very high resolution. Contrast this figure with the ideal spectrum shown in Figure 3.4. Figure from Ref. [21].

## 3.3 Spectrometry Results

### 3.3.1 Calibration using bench-top Gamma-ray Sources

$^{60}$Co, $^{137}$Cs, and $^{22}$Na were used to calibrate the $CeBr_3$ detector with their well-known γ-rays shown below in Figure 3.6. Each source was placed in front of the $CeBr_3$ detector for 30 minutes of irradiation at equal distances from the detector.

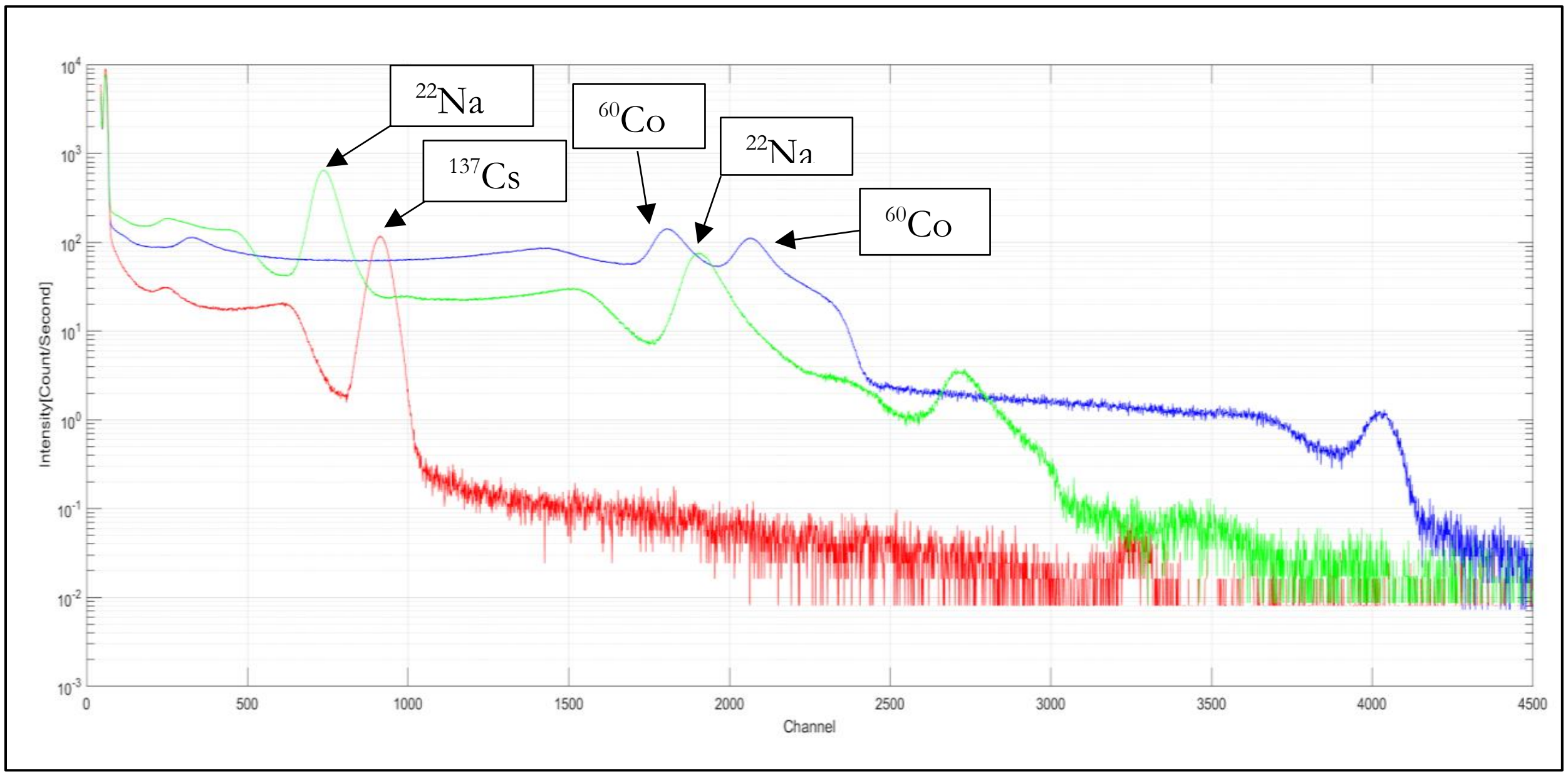

Figure 3.6: Spectra for $^{60}$Co, $^{137}$Cs, and $^{22}$Na.

Using a Gaussian fit we can determine each peak channel number and its standard deviation.

| Source | Gamma Ray 1 Energy (MeV) | Peak Channel [Ch.] | Standard deviation [Ch.] | Gamma Ray 2 Energy (MeV) | Peak Channel Number [Ch.] | Standard deviation [Ch.] |
|---|---|---|---|---|---|---|
| $^{60}$Co (Blue) | 1.17 | 1812.0 | 73.72 | 1.33 | 2070 | 82.88 |
| $^{137}$Cs (Red) | 0.662 | 913.3 | 39 | N/A | | |
| $^{22}$Na (Green) | 0.511 | 740.3 | 47.86 | 1.27 | 1914 | 79.33 |

Table 3.1: Gaussian Fit Results for the bench-top gamma-ray sources at 710 V.

We can then use the Peak Channel numbers and the known γ-ray energies to create a line of best fit for the energy calibration equation given by

$$E_\gamma[MeV] = E_\gamma[Channel] \times Gain\left[\frac{MeV}{Channel}\right] + constant[MeV]$$

where $E_\gamma[Channel]$ is the channel number of a point in the spectrum, $E_\gamma[MeV]$ is the corresponding energy at that point and $Gain\left[\frac{MeV}{Channel}\right]$ is a unique number of MeV per channel for each detector and applied voltage [19].

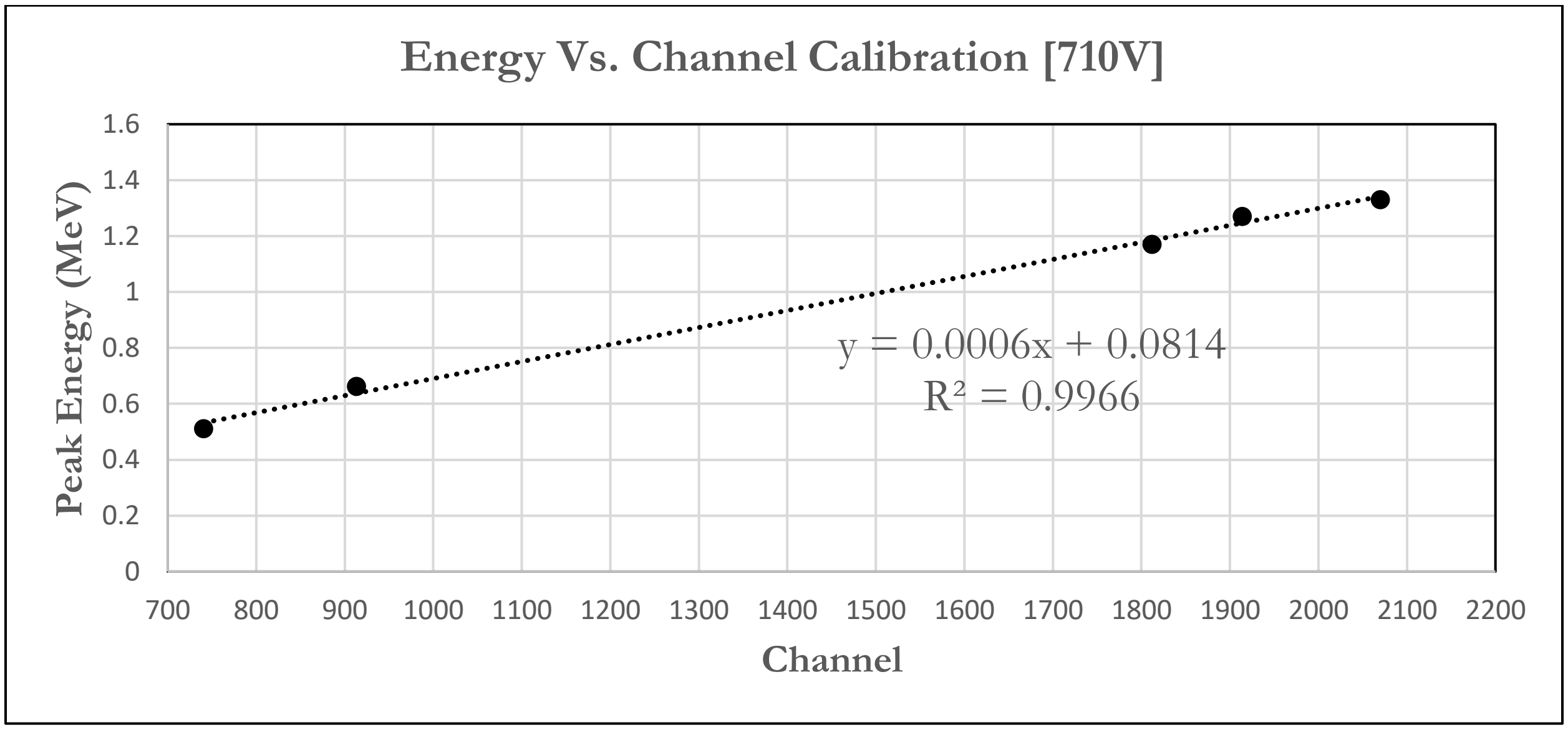

Figure 3.7: Linear fit to bench-top gamma-ray source data. Energy vs. Channel.

Figure 3.7 above shows that the detector behaves linearly at 710 V. The relationship between γ-ray energy and channel number is given by

$$E_\gamma[MeV] = E_\gamma[Channel]\times 0.0006\left[\frac{MeV}{Channel}\right] + 0.0814[MeV]$$

### 3.3.2 Calibration using High-Energy Sources

At 710 V, the detector appears initially to be linear. However, if we apply the fit equation on an AmBe spectrum covering the entire channel/energy range, the fit quickly becomes non-linear.

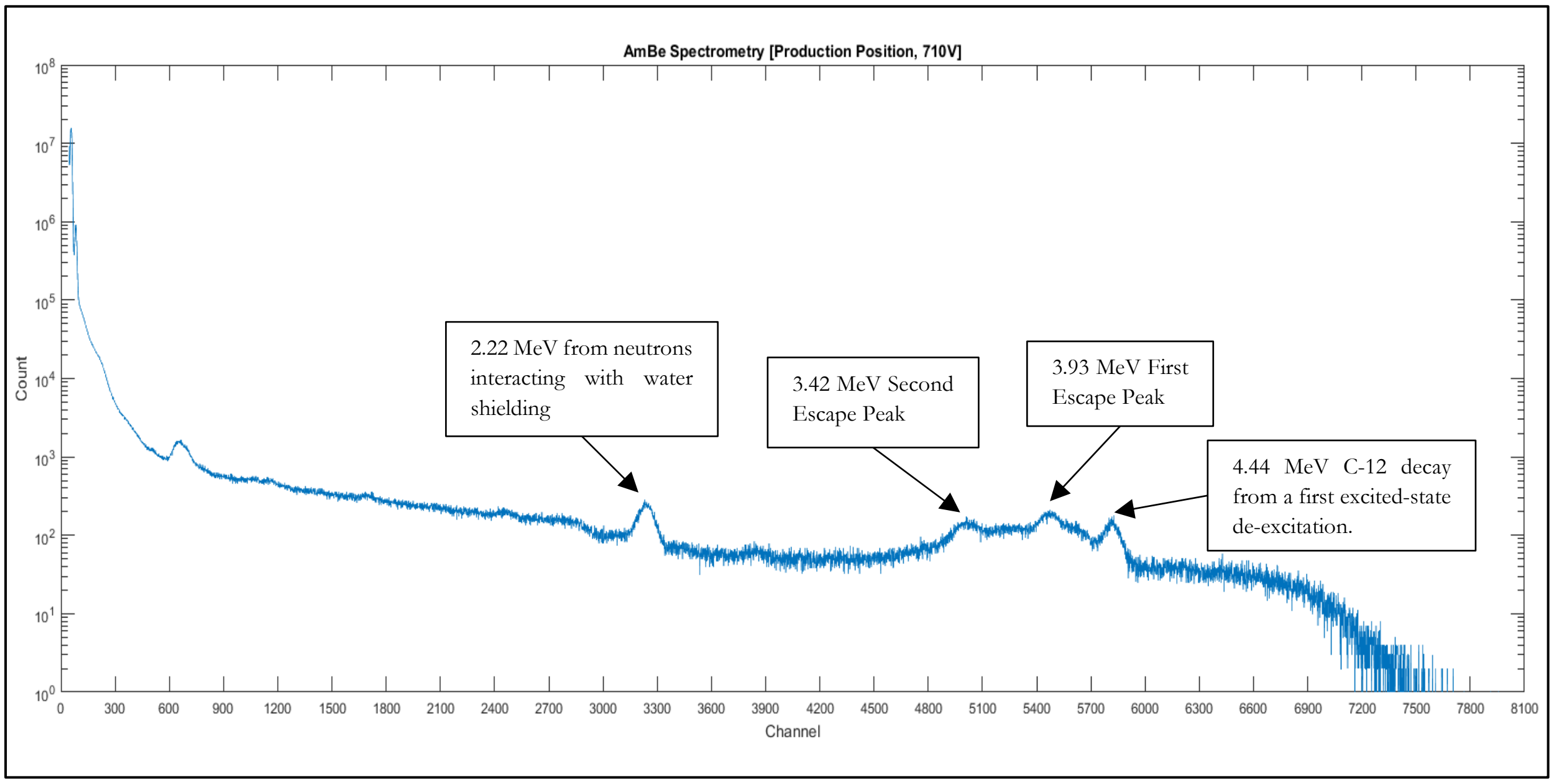

Figure 3.8: AmBe gamma-ray spectrum in production position at 710 V.

| Source | Expected Gamma Ray Energy (MeV) | Peak Channel [Gaussian fit] | E=Channel* 0.0006 +0.0814 (MeV) | Percent Error [%] |
|---|---|---|---|---|
| Second Escape Peak | 3.42 | 5026 | 3.0970 | 9.4444 |
| First Escape Peak | 3.93 | 5492 | 3.3766 | 14.0814 |
| $^{12}C$ (de-excitation of first-excited state) | 4.44 | 5824 | 3.5758 | 19.4639 |

Table 3.2: Linearity check at high energies. Due to the large uncertainties, we can conclude that the MCA was not performing linearly for the power supply settings of 710 V.

### 3.3.3 Calibration of Power Supply

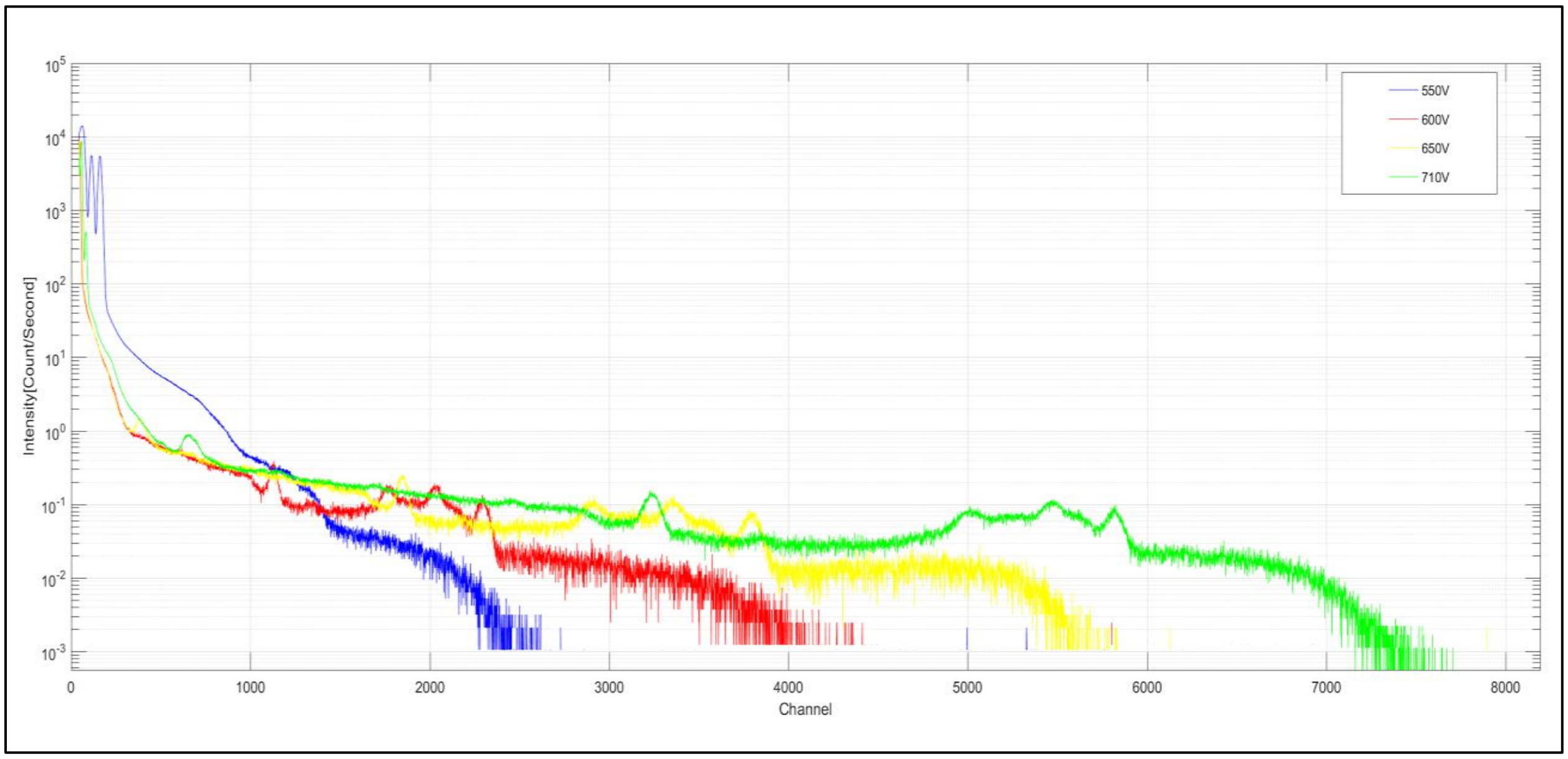

Figure 3.9: Gamma-Ray footprint of AmBe in production position using various power-supply settings.

Multiple measurements of the AmBe source in the production position were completed by varying the power-supply voltage to determine the optimal setting for linear operation of the detector. From inspection, we can see that the distance between the 4.44 MeV and first-escape peak is not equal to the distance between the first- and second- escape peaks when operating at 710 V which indicates non-linearly. However, even at 600 V where the distances are equal, there doesn't appear to be enough separation for the detection of the neutron capture by the B-10 present in the plastic blocks which were in the vicinity of the $CeBr_3$ detector. Therefore, 650 V is the best setting for further analysis. This is not ideal since it does not take advantage of the entire channel range available in the MCA8000D.

### 3.3.4 AmBe Source Spectrum at 650 V

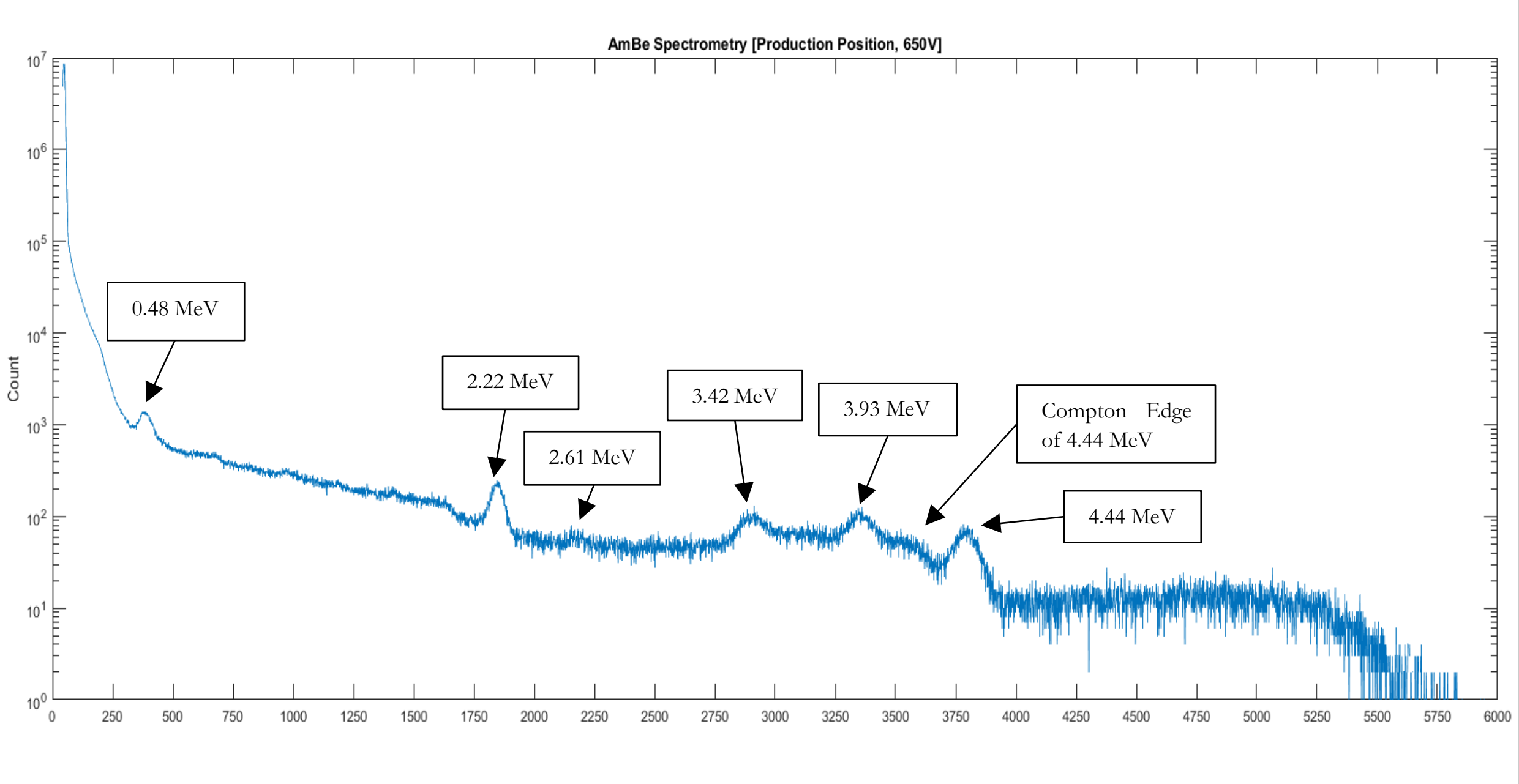

Figure 3.10: Gamma-Ray footprint of AmBe source in production position at 650 V.

| Peak Energy (MeV) | Source | Peak Channel [Ch.] | Standard Deviation (σ)[Ch.] | FWHM [2.35 σ] | Resolution [%] |
|---|---|---|---|---|---|
| 0.48 | Neutron capture by B-10 | 382.90 | 60.09 | 141.50 | 36.96 |
| 2.22 | Neutron capture by hydrogen in paraffin and water | 1845.00 | 49.06 | 115.53 | 6.26 |
| 2.61 | De-excitation of $^{208}$Pb (Th series) | 2175.00 | 131.30 | 309.19 | 14.22 |
| 3.42 | 4.44 MeV second-escape peak | 2911.00 | 127.60 | 300.46 | 10.32 |
| 3.93 | 4.44 MeV first-escape peak | 3355.00 | 111.50 | 262.56 | 7.83 |
| 4.44 | De-excitation of $^{12}C^{*}$ | 3791.00 | 88.43 | 208.24 | 5.49 |

Table 3.3: Sources of gamma-ray peaks labeled in Figure 3.10.

It is generally expected that the detector resolution will improve with an increase of peak energy. However, that is not fully consistent with the results found in Table 3.3. The average energy resolution is 13.51% which is far greater than the 3.8% advertised by the manufacturer (see Appendix E). One reason for such a discrepancy is due to the STF source being a large, ill-defined volume rather than a point source which is used by the manufacturer. The linear fit of the peaks produced at 650 V was much more linear which included a larger variance of energy than the fit in Figure 3.7.

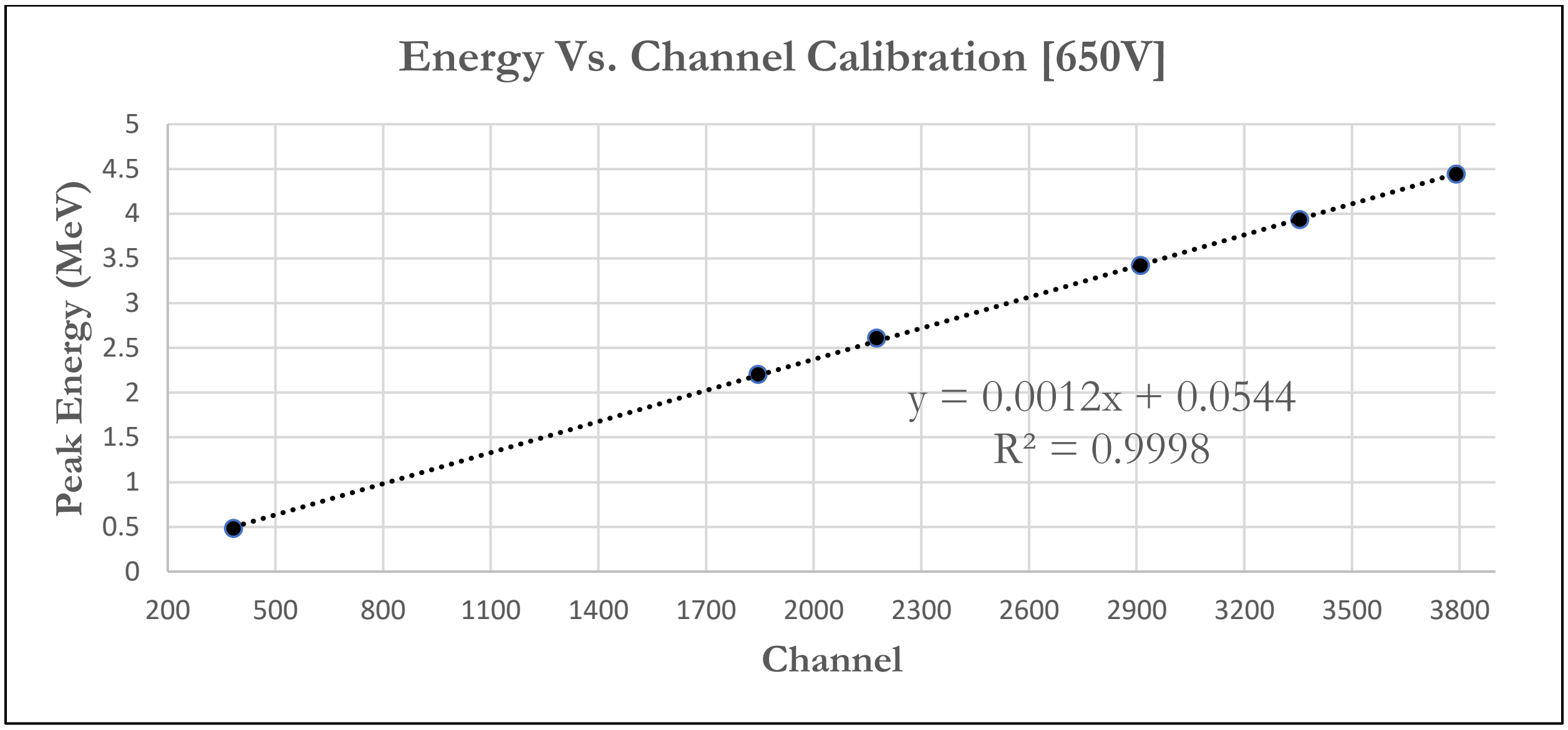

Figure 3.11: Linear fit of detected gamma-ray peaks detailed in Table 3.3.

### 3.3.5 Compton Edge of AmBe Source at 650 V

The Compton-edge energy corresponding to the 4.44 MeV peak associated with the de-excitation of $^{12}C$ can be determined using the equation in section 3.2.1 as 4.2 MeV. There are two common methods to estimate the channel location of the 4.2 MeV peak in the measured spectra. According to Flynn, the Compton edge is located 4% below 50% of the height of the peak whereas Knox claims that the Compton edge is at 89% of the height of the peak [19].

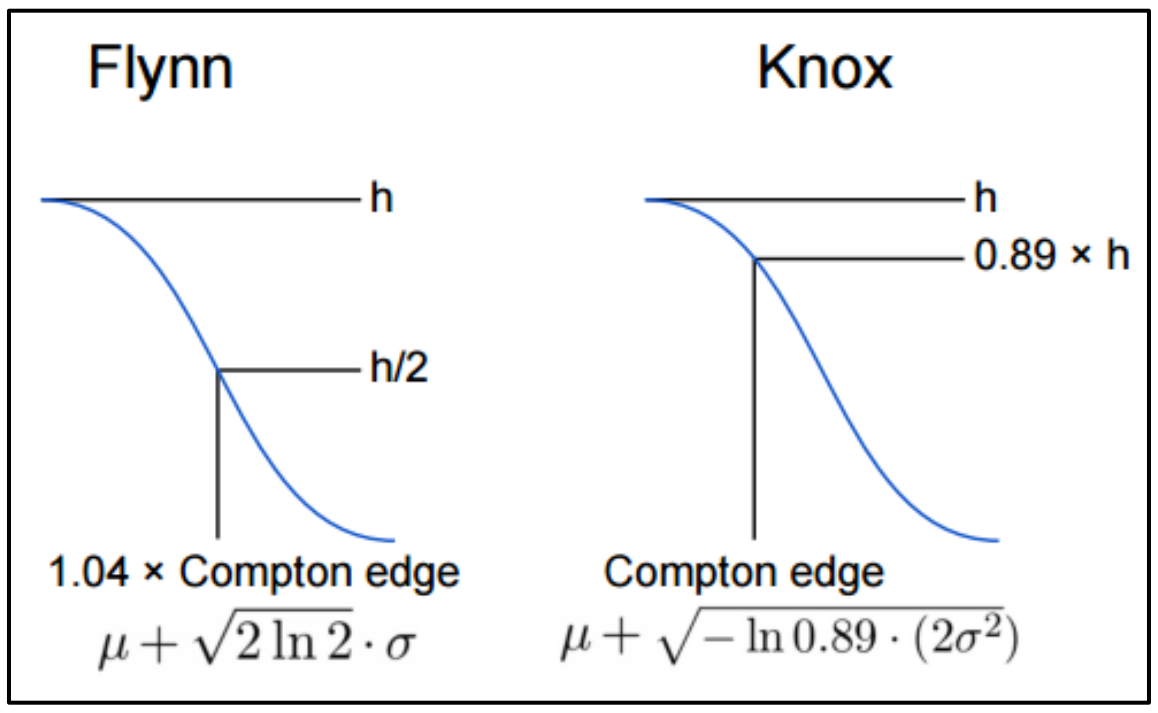

Figure 3.12: Illustration of the Flynn and Knox methods. These methods are used in calculating the location of a Compton edge. Figure from Ref. [19].

| Peak Energy (MeV) | Source | Standard Deviation [Ch.] | Compton Edge Channel [Gaussian fit] | Compton Edge Channel [Flynn] | Compton Edge Channel [Knox] |
|---|---|---|---|---|---|
| 4.2 | Compton Edge 4.44 MeV | 188.5 | 3518.0 | 3961.9 | 3609.0 |

Table 3.4: Compton-edge results obtained using the Gaussian, Flynn and Knox methods.

New linear fits were obtained by including each of the Compton edge channel numbers in Table 3.4 with the photopeak data in Table 3.3. The largest coefficient of determination was found using the Knox method to be 0.998. This agrees with the observations of Scherzinger [22].

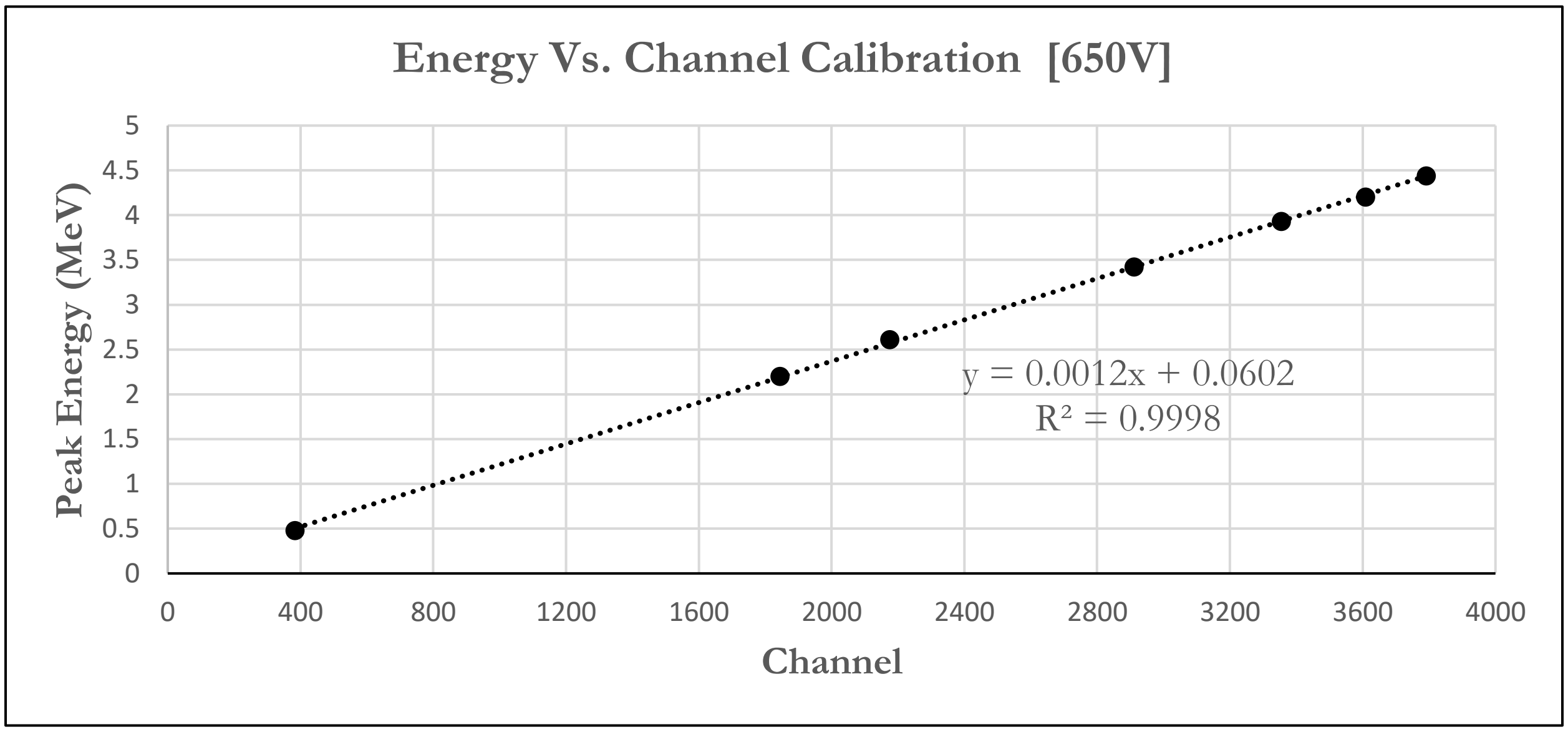

Figure 3.13: Linear fit of the source peaks in Table 3.3 including the 4.44 MeV Compton-edge determined using the Knox Method.

### 3.3.6 Tank Shielding

The γ-ray footprint of the AmBe neutron source was measured using the $CeBr_3$ detector at various positions within the tank for 30 minutes each. The AmBe source was moved to the park position, production position and top position. The $CeBr_3$ detector was placed 10 cm from the outside wall of the tank for a total horizontal distance of 39.5 cm away from the AmBe source.

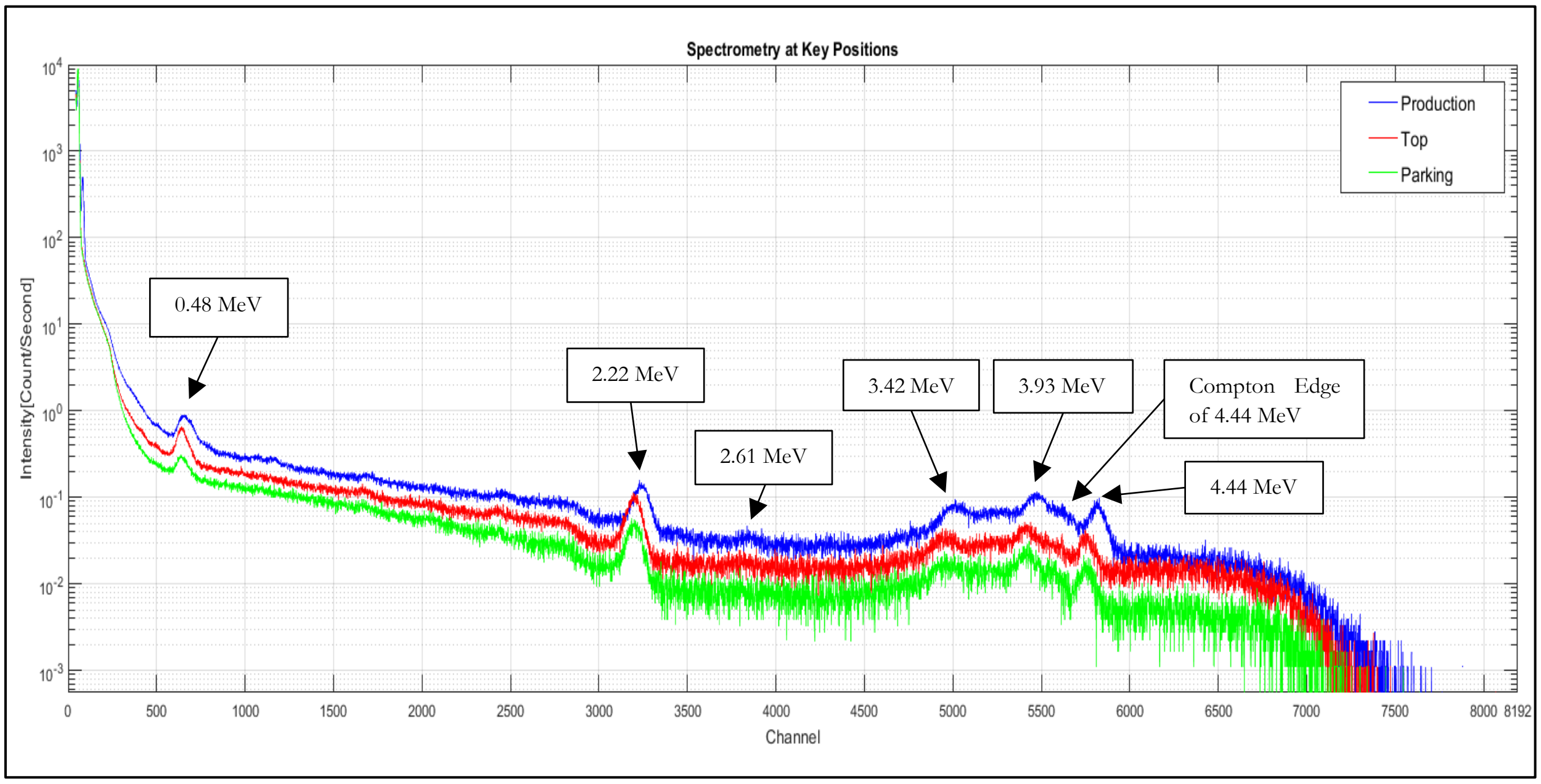

Figure 3.14: AmBe Gamma-Ray footprint at the production, top and parking positions.

| Source Position | Polyethylene Plug | Total Count Rate from channel 500 to 8000 [Hz] |
|---|---|---|
| Production | No | 705.1 |
| Top | No | 419.5 |
| Park | No | 253.8 |

Table 3.5: Gamma-ray rates for various configurations.

Table 3.5 shows the count rates measured for the configurations by integrating the count rate over the γ-ray range. This shows that the park position reduces the γ-ray rate in the $CeBr_3$ detector by a factor of 3.

# Summary

The purpose of this project was to further investigate the radiological footprint of an AmBe source within the source tank in the Source-Testing Facility at the Department of Nuclear Physics at Lund University. Spectra and rates from various positions within the tank were needed to serve as benchmarks for precisely simulating the γ-ray fields associated with the AmBe source. This was completed using a $CeBr_3$ γ-ray detector. Signals from the detector were read and histogrammed using a Multi-Channel Analyzer. The resulting spectra was then used to verify the resolution of the detector.

To enable the mapping of various relative source and detector configurations, design and implementation of a precise radioactive Source-positioning System was required. The goal of the Source-positioning System was also to enable the absolute characterization of the neutron and γ-ray field from the Detector-Development Platform such that the reproducibility of the measurements can then be established.

Results indicate that the park position which is practically an off position for the source reduces the γ-ray rate in the $CeBr_3$ detector by a factor of 3.

# Appendix A: Binding energy calculation for $^{241}$Am

**Constants:**

$Mass\ of\ {}^{241}_{95}Am = 241.05682\ u$
$Mass\ of\ e^{-} = 0.000549\ u$
$Mass\ of\ n^{0} = 1.008665\ u$
$Mass\ of\ p^{+} = 1.0072765\ u$

**Mass Defect Calculation:**

$Atomic\ mass\ of\ {}^{241}_{95}Am = m_{{}^{241}_{95}Am} - 95m_{e^-} = 241.004665\ u$
$Atomic\ mass\ of\ all\ nucleons = 146n^{0} + 95p^{+} = 242.956367\ u$
$Mass\ Defect = 242.956367 - 241.004665 = 1.951692\ u$

**Binding Energy Calculation:**

$E = mc^2$
$1\ amu = 931.502\ MeV$
$Binding\ Energy\ of\ {}^{241}_{95}Am = 1.951692 \times 931.502$

$$Binding\ Energy\ of\ {}^{241}_{95}Am\ per\ nucleon = \frac{Binding\ Energy}{241} = 7.54359\ MeV$$

## Appendix B: Ground State Q-value calculation for $^{241}$Am decay

**Constants:**

$Mass\ of\ ^{241}_{95}Am = 241.05682\ u$
$Mass\ of\ ^{237}_{93}Np = 237.04817\ u$
$Mass\ of$ α particle $= 4.001506\ u$
$Mass\ of\ e^{-} = 0.000549\ u$

**Mass Calculations:**

$Atomic\ mass\ of\ ^{241}_{95}Am = m_{^{241}_{95}Am} - 95m_{e^-} = 241.004665\ u$
$Atomic\ mass\ of\ ^{237}_{93}Np = m_{^{237}_{93}Am} - 93m_{e^-} = 236.9971108\ u$

**Q-value Calculation:**

$Q_1 = (241.004665 - 4.001506 - 236.9971108)\ (c^2)u = 0.0060482\ (c^2)u$

$Q_1 = (0.0060482\ (c^2)u)(^{931.502\ \text{MeV}}/_{(c^2)u} = 5.6339103964\ \text{MeV}$

## Appendix C: Ground State Q-value calculation of carbon reaction

(Assuming no kinetic energy in α-particle. This is the case when all of $Q_1$ is transferred to $^{237}$Np due to a direct ground state decay from $^{241}$Am.)

**Constants:**

$Mass\ of\ {}^{12}_{6}C = 12u$

$Mass\ of\ n^0 = 1.008665\ u$

$Mass\ of\ {}^{9}_{4}Be = 9.0121821$ u

$Mass\ of\ \alpha$ particle $= 4.001506\ u$

$Mass\ of\ e^- = 0.000549\ u$

**Mass Calculations:**

$Atomic\ mass\ of\ {}^{12}_{6}C = m_{{}^{12}_{6}C} - 6m_{e^-} = 11.996706$ u

$Atomic\ mass\ of\ {}^{9}_{4}Be = m_{{}^{9}_{4}Be} - 4m_{e^-} = 9.0099861$ u

**Q-Value Calculation:**

$Q_2 = (9.0099861 + 4.001506 - 11.996706 - 1.008665)\ (c^2)u = 0.0061211(c^2)u$

$Q_2 = (0.0061211\ (c^2)u)(^{931.502\ \text{MeV}}/_{(c^2)u} = 5.7018168922\ \text{MeV}$

# Appendix D: SPS Parts

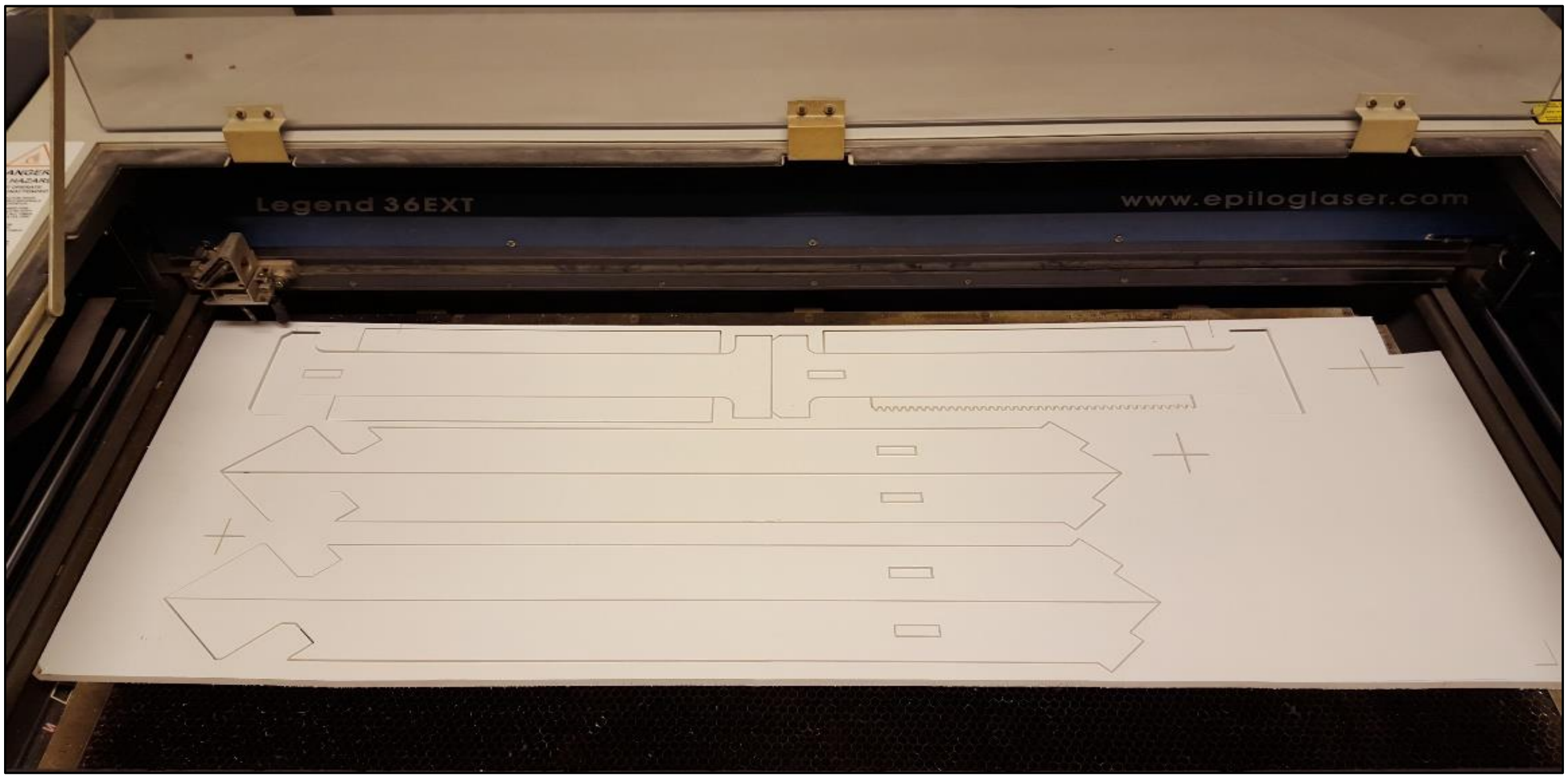

**Plexiglas Board**

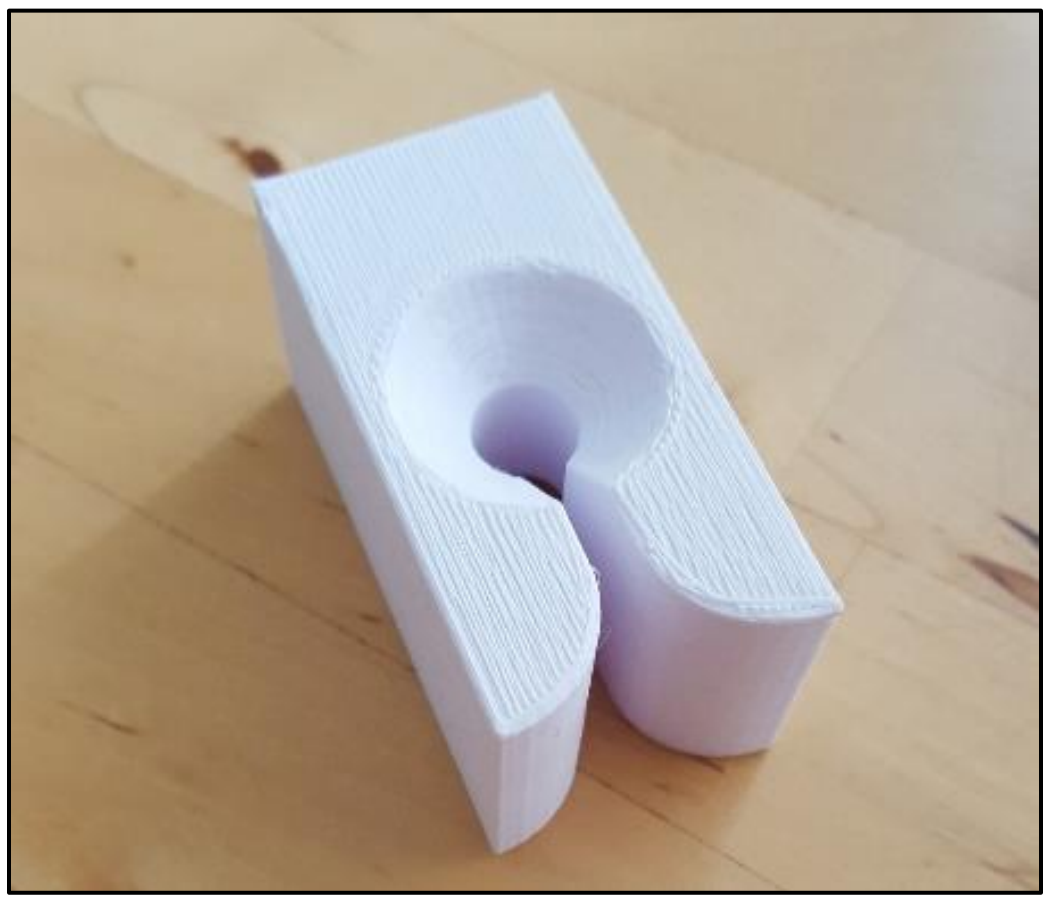

**Custom Gate 3D-printed**

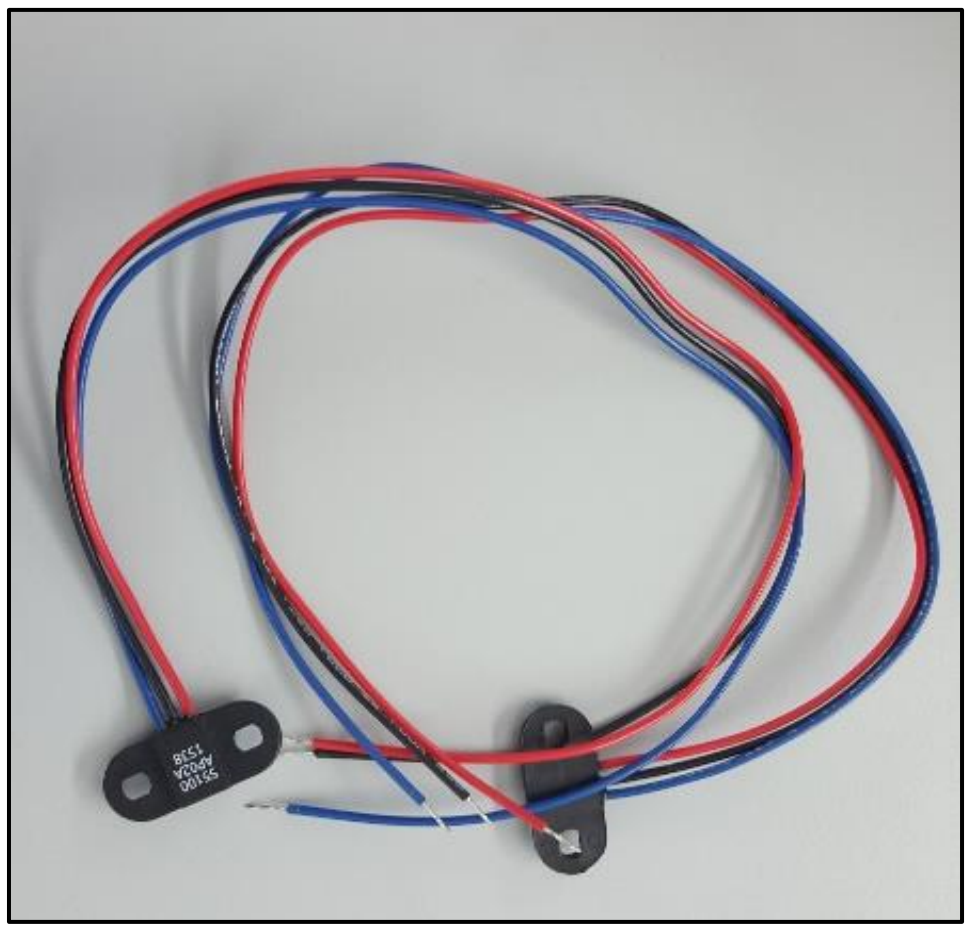

**Hamlin 55100 AP 02 A Hall Effect Sensors**

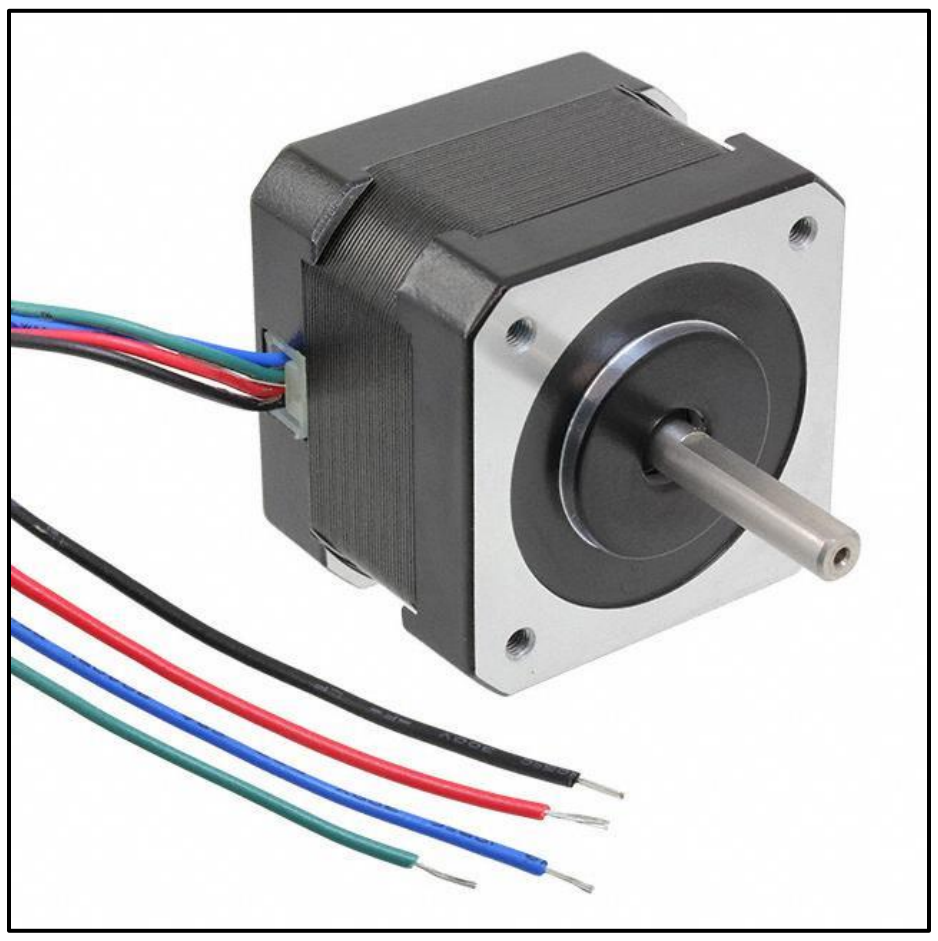

**2 Trinamic QSH4218-35-10-027 Stepper Motors, BiPolar, 27 N-cm, 1 A, 5.3 ohm, 6.6 mH**

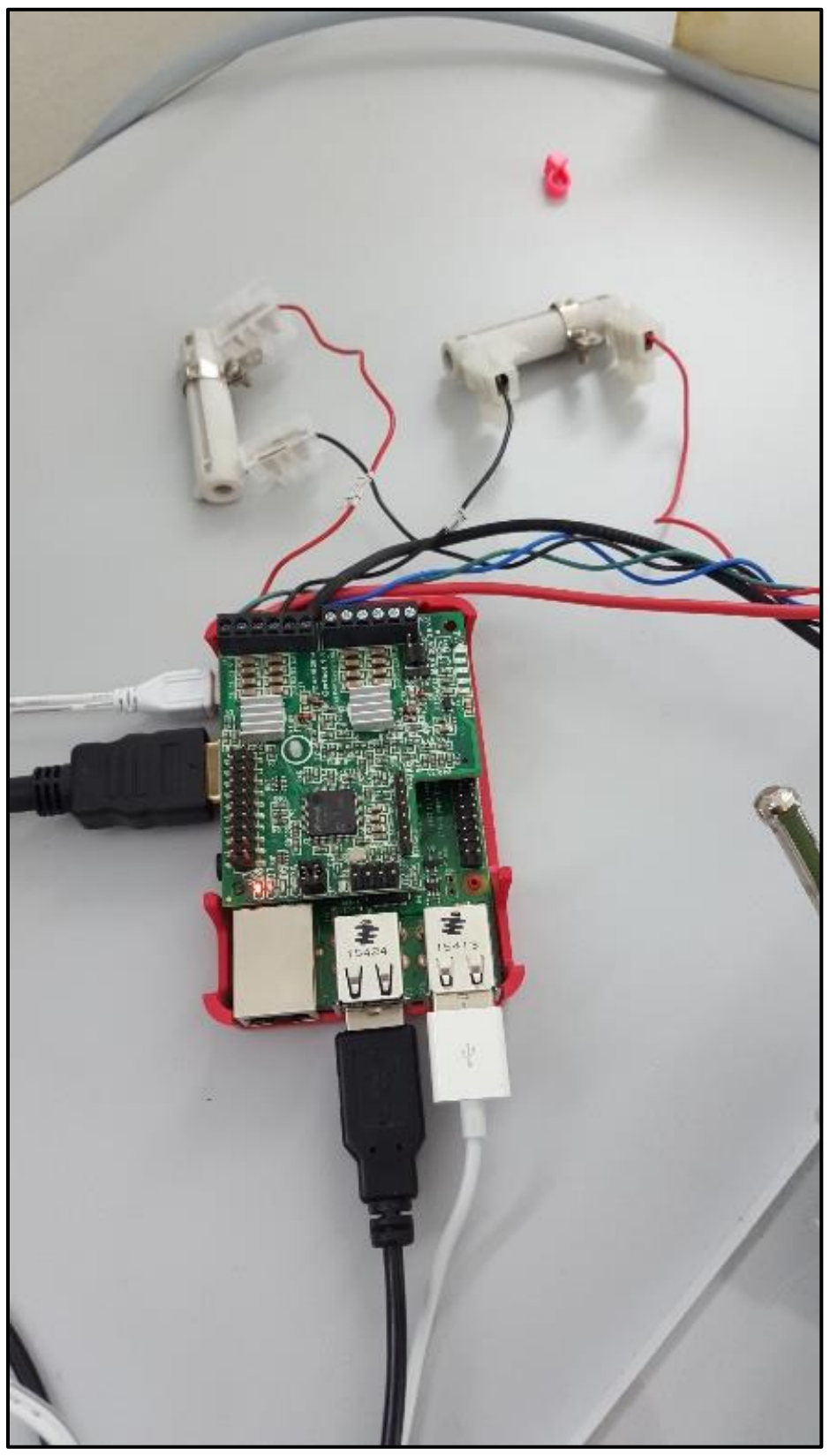

**Raspberry Pi3 and GertBoard GertBot**

# Appendix E: Power Supply Tests & Detector Data Sheet

| Vmon | Imon[uA] | Trip | Vset | Iset[uA] | Ramp Up | Ramp Down |
|---|---|---|---|---|---|---|
| 634 | 109.6 | Yes | 710 | 110 | 10 | 50 |
| 634.2 | 109.6 | Yes | 710 | 110 | 50 | 50 |
| 682.4 | 118 | Yes | 710 | 118.33 | 50 | 50 |
| 709.6 | 122.65 | Yes | 710 | 122.067 | 50 | 50 |
| 709.6 | 122.65 | Yes | 710 | 122.1 | 50 | 50 |
| 709.6 | 122.7 | No | 710 | 123 | 50 | 50 |
| 709.8 | 122.7 | No | 710 | 130 | 50 | 50 |
| 709.8 | 122.7 | No | 710 | 140 | 50 | 50 |
| 710.6 | 122.85 | No | 710 | 140 | 50 | 50 |

From Paul Schotanus <paul@scionix.nl>

"The datasheet I have is attached . *The current drawn by the detector is totally the voltage divder design. This is NOT the photocurrent !* If you are running the detector at a high count rate the current should be suffcient to keep the amplifcation constant.

Our standard Voltage Divider has a resistor value of 6 Mohm. This means that at 1000 V the bleeder current is 0.16 mA.
In fact the voltage on such a detector should be approx. 800V I guess for a resaible output. You should look at the linearity of the spectrum with e.g. Cs-137 and Co-60. If your voltage is too high, the PMT starts to become alinear.

Best regards

Paul"

SCIONIX
Dedicated Scintillation Detectors

# T E S T S H E E T

| | | |
|---|---|---|
| **Detector** | : | 38B38/2ME1CEBRX6NEG |
| **Crystal** | : | CeBr3 |
| **Readout** | : | Hamamatsu 2" Type R6231 |
| **Serial Number** | : | SFU649 |

max current
0.1 mA
=
100 μA
=

430 V = 75 μA
489 V = 85 μA
550 V = 95 μA
518 V => 90 μA
ramp up doesn't matter

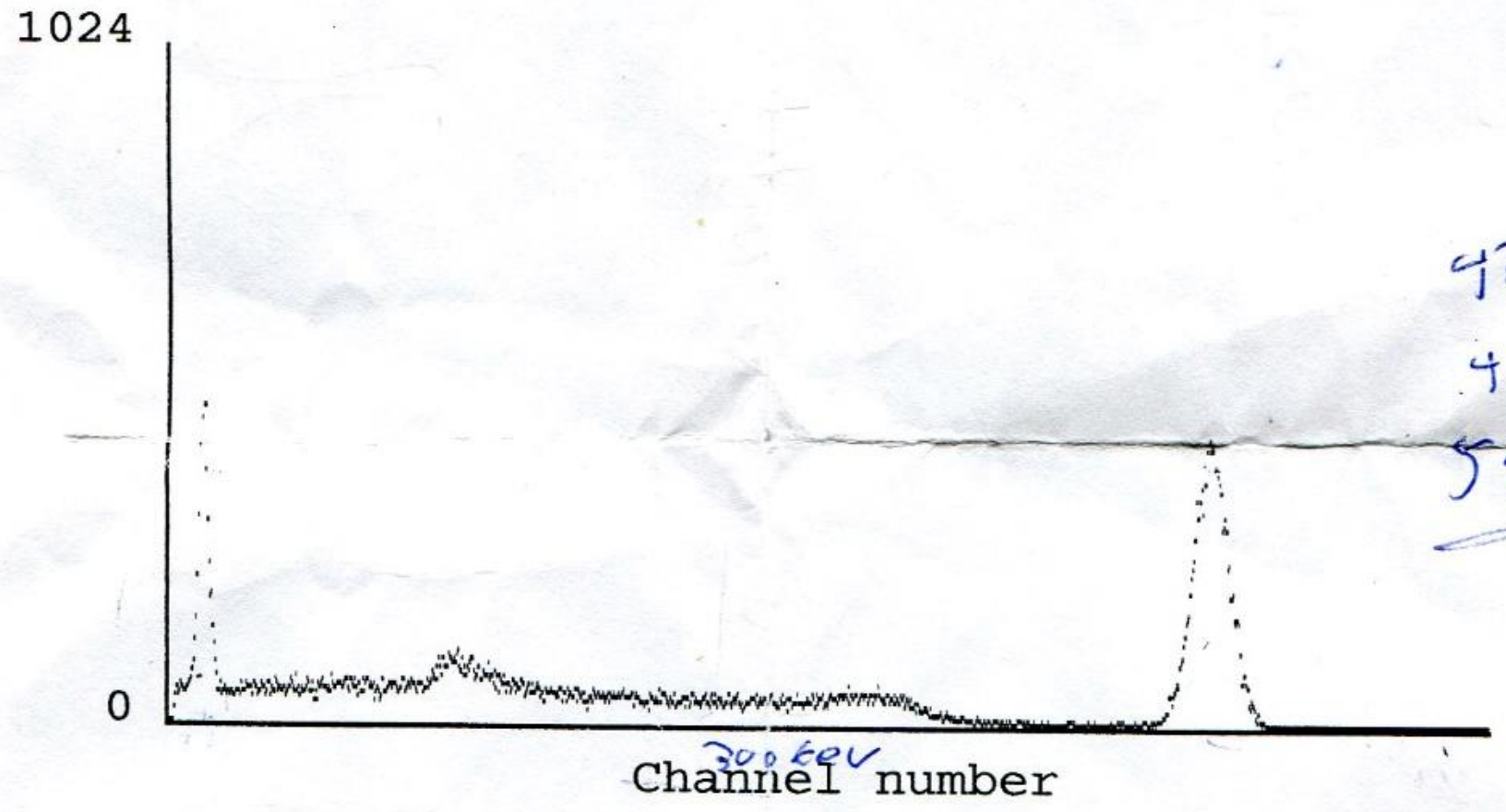

# T E S T  R E S U L T S

| | | |
|---|---|---|
| **Nuclide** | : | Cs-137 (662 keV) |
| **Energy Resolution** | : | 3.8 % |
| **Noise** | : | N/A |
| **Peak to Valley** | : | N/A |
| **Am241 or Cf252 GEE** | : | N/A |
| **High Voltage** | : | -710 V |
| **Date** | : | 01 April 2016 |
| **Tested by** | : | Erwin Bodewits ........ |

SCIONIX HOLLAND BV, P.O. Box 143, 3980 CC Bunnik, The Netherlands
Tel. +31 (0)30-65 70312, Fax +31 (0)30-65 67563, Web: www.scionix.nl, E-mail: purchasing@scionix.nl

# Appendix F: Matlab Scripts

### Gaussian Fit

```
function [fitresult, gof] = createFit(CEch,CE);

importdata('volt calibration.mat');

[xData, yData] = prepareCurveData(CEch,CE);


ft = fittype( 'gauss1' );
opts = fitoptions( 'Method', 'NonlinearLeastSquares' );
opts.Display = 'Off';
opts.Lower = [0 3477 0 ];
opts.StartPoint = [50 3500 45.45 ];
opts.Upper = [2000 3667 400];

% Fit model to data.
[fitresult, gof] = fit( xData, yData, ft, opts );

% Plot fit with data.
semilogy(xData, yData);
hold on;
axis([3000 4000 0 2000]); %xmin xmax ymin ymax
h = plot( fitresult, xData, yData );
legend( h, 'CeBr3 vs. channel', 'keV', 'Location', 'NorthEast' );
% Label axes
xlabel Channel
ylabel Peak
grid on
```

### Graphing by Intensity

```
A=production
B=top
C=parking

production_intensity=A./1800
top_intensity=B./1800
parking_intensity=C./1800

semilogy(channel,production_intensity,'blue')
hold on
semilogy(channel,top_intensity,'red')
hold on
semilogy(channel,parking_intensity,'green')
hold on


axis([0 8192 0 10000])
grid on
```